\documentclass[12pt]{article}%
\usepackage{heck}
\usepackage{graphicx}
\usepackage{makeidx}
\usepackage{multicol}
\usepackage{geometry}
\usepackage{amsfonts}
\usepackage{amssymb}
\usepackage{amsmath}%
\begin{document}

\date{October, 2006}

\preprint{hep-th/0610005 \\ HUTP-06/A0039}

\institution{HarvardU}{Jefferson Physical Laboratory, Harvard University, Cambridge,
MA 02138, USA}%

%TCIMACRO{\TeXButton{Institution}{\institution{MIT}%
%{Center for Theoretical Physics, MIT, Cambridge, MA 02139, USA}}}%
%BeginExpansion
\institution{MIT}%
{Center for Theoretical Physics, MIT, Cambridge, MA 02139, USA}%
%EndExpansion
%

%TCIMACRO{\TeXButton{Title}{\title{Crystal Melting and Black Holes}}}%
%BeginExpansion
\title{Crystal Melting and Black Holes}%
%EndExpansion
%

%TCIMACRO{\TeXButton{Authors}{\authors{Jonathan J. Heckman\worksat{\HarvardU
%,}\footnote{e-mail: {\tt jheckman@fas.harvard.edu}}
%and Cumrun Vafa\worksat{\HarvardU,\MIT}\footnote{e-mail: {\tt
%vafa@string.harvard.edu}}}}}%
%BeginExpansion
\authors{Jonathan J. Heckman\worksat{\HarvardU,}\footnote{e-mail: {\tt
jheckman@fas.harvard.edu}}
and Cumrun Vafa\worksat{\HarvardU,\MIT}\footnote{e-mail: {\tt
vafa@string.harvard.edu}}}%
%EndExpansion

\abstract{It has recently been shown that the statistical mechanics of crystal melting
maps to A-model topological string amplitudes on non-compact Calabi-Yau
spaces.  In this note we establish a one to one correspondence between two
and three dimensional crystal melting configurations and certain BPS black holes given by branes wrapping
collapsed cycles on the orbifolds $\mathbb{C}^2/\mathbb{Z}_{n}$ and
$\mathbb{C}^3/\mathbb{Z}_{n}\times\mathbb{Z}_{n}$ in the large $n$ limit.  The
ranks of gauge groups in the associated gauged quiver quantum mechanics determine
the profiles of crystal melting configurations and the process of melting maps to flop transitions which leave
the background Calabi-Yau invariant.  We explain the connection between these two realizations
of crystal melting and speculate on the underlying physical meaning.}%

%TCIMACRO{\TeXButton{Maketitle}{\maketitle}}%
%BeginExpansion
\maketitle
%EndExpansion

\section{Introduction}

An important feature of string theory is how geometric data in a string
compactification appear in the associated low energy effective theory. \ A
striking example is provided by space-time filling brane probes of orbifold
singularities. \ From the perspective of the quiver gauge theory, a
topology-changing flop transition appears as a Seiberg-like duality which
alters the ranks, matter content, and interaction terms of the gauge theory
\cite{CFIKV,DouglasBerenstein}. \ It is therefore of interest to study the
effect of flop transitions on more general brane configurations.

A related issue concerns what degrees of freedom in string theory replace
macroscopic notions of classical geometry at small distance scales. \ In
\cite{QuantumFoam} it was shown that the target space formulation of the
topological A-model on $%
%TCIMACRO{\U{2102} }%
%BeginExpansion
\mathbb{C}
%EndExpansion
^{3}$ describes quantum foam on the same geometry. \ Remarkably, the partition
sum of this model is identical to that of three dimensional crystal melting
\cite{CrystalMelting}. \ In two dimensions, crystal melting adds squares to a
2d Young tableau. \ In three dimensions, crystal melting stacks cubes at the
corner of a room. \ It is believed that other crystal melting models determine
A-model amplitudes on more general non-compact Calabi-Yau spaces.

Topological strings have also recently appeared in the physics of four
dimensional black holes. \ The work of \cite{OSV} presented evidence that the
indexed partition function for a mixed ensemble of four dimensional Calabi-Yau
black holes with fixed magnetic charges and electric chemical potentials is
determined by topological string theory on the same Calabi-Yau space. \ This
suggests that other ensembles of black holes are also of relevance.

In this note we combine these themes and show that the crystal melting
configurations of topological string theory are in one to one correspondence
with certain BPS\ black holes given by branes wrapped on collapsed cycles in
the supersymmetric orbifolds $%
%TCIMACRO{\U{2102} }%
%BeginExpansion
\mathbb{C}
%EndExpansion
^{2}/%
%TCIMACRO{\U{2124} }%
%BeginExpansion
\mathbb{Z}
%EndExpansion
_{n}$ and $%
%TCIMACRO{\U{2102} }%
%BeginExpansion
\mathbb{C}
%EndExpansion
^{3}/%
%TCIMACRO{\U{2124} }%
%BeginExpansion
\mathbb{Z}
%EndExpansion
_{n}\times%
%TCIMACRO{\U{2124} }%
%BeginExpansion
\mathbb{Z}
%EndExpansion
_{n}$ in the limit of large $n$. \ The black holes in this spectrum are
generated by all possible flop transitions which leave the background
Calabi-Yau geometry invariant, and the stability of these configurations is
determined by the rigid structure of the partially melted crystal.
\ Additionally, in both cases the duality group generated by these transitions
corresponds to the Weyl group of an infinite dimensional algebra. \ The
character of the basic representation of each algebra coincides with the
corresponding partition sum over black hole charges. \ We believe that this
technique of generating BPS\ spectra on a Calabi-Yau space using geometry
preserving flop transitions may also be of use in a wider context than that
presented here.

At low energies, these charge configurations are well-described by gauged
quiver quantum mechanics. \ The ranks of the gauge group factors determine D0-
and D2-branes wrapped on collapsed cycles in $%
%TCIMACRO{\U{2102} }%
%BeginExpansion
\mathbb{C}
%EndExpansion
^{2}/%
%TCIMACRO{\U{2124} }%
%BeginExpansion
\mathbb{Z}
%EndExpansion
_{n}$ and D0-, D2- and D4-branes wrapped on collapsed cycles in $%
%TCIMACRO{\U{2102} }%
%BeginExpansion
\mathbb{C}
%EndExpansion
^{3}/%
%TCIMACRO{\U{2124} }%
%BeginExpansion
\mathbb{Z}
%EndExpansion
_{n}\times%
%TCIMACRO{\U{2124} }%
%BeginExpansion
\mathbb{Z}
%EndExpansion
_{n}$. \ In the limit of large charges, a stable configuration in $%
%TCIMACRO{\U{2102} }%
%BeginExpansion
\mathbb{C}
%EndExpansion
^{2}/%
%TCIMACRO{\U{2124} }%
%BeginExpansion
\mathbb{Z}
%EndExpansion
_{n}$ and $%
%TCIMACRO{\U{2102} }%
%BeginExpansion
\mathbb{C}
%EndExpansion
^{3}/%
%TCIMACRO{\U{2124} }%
%BeginExpansion
\mathbb{Z}
%EndExpansion
_{n}\times%
%TCIMACRO{\U{2124} }%
%BeginExpansion
\mathbb{Z}
%EndExpansion
_{n}$ will produce a local model for a BPS\ black hole in six and four
dimensions, respectively. \ From the perspective of the effective theory, a
flop transition which leaves the classical geometry invariant may alter the
ranks of the gauge groups but will always preserve both the adjacency of
bifundamentals in the quiver as well as the form of the superpotential.

In order to describe the precise action of flop transitions on $%
%TCIMACRO{\U{2102} }%
%BeginExpansion
\mathbb{C}
%EndExpansion
^{3}/%
%TCIMACRO{\U{2124} }%
%BeginExpansion
\mathbb{Z}
%EndExpansion
_{n}\times%
%TCIMACRO{\U{2124} }%
%BeginExpansion
\mathbb{Z}
%EndExpansion
_{n}$, we will find it necessary to treat the fractional branes of the
orbifold theory as objects in the bounded derived category of coherent
sheaves. \ Each basis of fractional branes is generated by a series of
mutations on an exceptional collection of sheaves supported on a complex
surface of the resolved geometry. \ Further, the exceptional collections
appropriate for describing three dimensional crystal melting are all supported
on a complex surface in the canonical smooth resolution of $%
%TCIMACRO{\U{2102} }%
%BeginExpansion
\mathbb{C}
%EndExpansion
^{3}/%
%TCIMACRO{\U{2124} }%
%BeginExpansion
\mathbb{Z}
%EndExpansion
_{n}\times%
%TCIMACRO{\U{2124} }%
%BeginExpansion
\mathbb{Z}
%EndExpansion
_{n}$.

Exceptional collections are most commonly used in the study of type
IIB\ space-time filling D3-brane probes of local Calabi-Yau singularities
\cite{CFIKV,DouglasBerenstein,HerzogSeiberg,AspinwallTilting,WijnholtLargeVol}%
. \ We note that there are typically many exceptional collections which
generate the \textit{same} four dimensional quiver gauge theory. \ This
suggests the presence of a large gauge symmetry or redundancy in the
description of four dimensional quiver gauge theories in terms of exceptional
sheaves. \ Indeed, in the case of four dimensional gauge theories, gauge
anomaly considerations require that the rank assignments remain unchanged
under flop transitions which preserve the classical geometry. \ A perhaps
surprising result of this note is that in the more general case of BPS\ black
holes described by gauged quiver quantum mechanics, different exceptional
collections \textit{will} produce different rank assignments in the quiver
theory and will therefore break the gauge symmetry described above.

In our setup, the profile or \textquotedblleft height\textquotedblright\ of
the partially melted crystal is encoded in the ranks of gauge groups in the
quiver theory. In some sense, this is to be expected. \ The target space
formulation of topological string theory in terms of crystal melting in two
and three dimensions corresponds to counting D0-brane bound states in the
topologically twisted $U(1)$ gauge theories of a D4-brane and D6-brane filling
$%
%TCIMACRO{\U{2102} }%
%BeginExpansion
\mathbb{C}
%EndExpansion
^{2}$ and $%
%TCIMACRO{\U{2102} }%
%BeginExpansion
\mathbb{C}
%EndExpansion
^{3}$, respectively. \ Since a D0-brane in the orbifolds $%
%TCIMACRO{\U{2102} }%
%BeginExpansion
\mathbb{C}
%EndExpansion
^{2}/%
%TCIMACRO{\U{2124} }%
%BeginExpansion
\mathbb{Z}
%EndExpansion
_{n}$ and $%
%TCIMACRO{\U{2102} }%
%BeginExpansion
\mathbb{C}
%EndExpansion
^{3}/%
%TCIMACRO{\U{2124} }%
%BeginExpansion
\mathbb{Z}
%EndExpansion
_{n}\times%
%TCIMACRO{\U{2124} }%
%BeginExpansion
\mathbb{Z}
%EndExpansion
_{n}$ lifts to $n$ and $n^{2}$ image branes in $%
%TCIMACRO{\U{2102} }%
%BeginExpansion
\mathbb{C}
%EndExpansion
^{2}$ and $%
%TCIMACRO{\U{2102} }%
%BeginExpansion
\mathbb{C}
%EndExpansion
^{3}$ respectively, the untwisted RR D0-brane charge of the orbifold theory is
up to a factor of $1/n$ and $1/n^{2}$ this sum in ranks. \ We now describe in
more detail these crystal melting configurations.

\begin{figure}
[ptb]
\begin{center}
\includegraphics[
height=1.4687in,
width=2.5396in
]%
{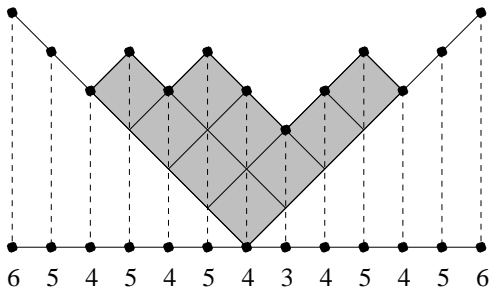}%
\caption{A 2d Young tableau defines a height function on the one dimensional
lattice given by the integers. \ The numbers shown indicate height assignments
for the given tableau. \ These heights are the rank assignments in the large
$n$ $\mathbb{C} ^{2}/\mathbb{Z} _{n}$ quiver theory.}%
\label{newheightexample}%
\end{center}
\end{figure}

As mentioned, the \textquotedblleft height function\textquotedblright\ of the
crystal specifies the ranks in the quiver theory. \ For 2d crystals, this
integer valued function satisfies:%
\begin{equation}
h(i)-h(i+1)=\pm1 \label{derivative}%
\end{equation}
for all integers $i$. \ The profile of this function is given by rotating the
Young tableau by $135^{\circ}$. \ See figure (\ref{newheightexample}) for an
example. \ To describe heights for 3d crystals, we use dimer models. \ Recent
reviews on the mathematical physics of dimer models are given in
\cite{KenyonDimerReview,KOSDimerReview}.%
%TCIMACRO{\FRAME{ftbpFU}{2.5396in}{1.4687in}{0pt}{\Qcb{A 2d Young tableau
%defines a height function on the one dimensional lattice given by the
%integers. \ The numbers shown indicate height assignments for the given
%tableau. \ These heights are the rank assignments in the large $n$ $\U{2102}
%^{2}/\U{2124} _{n}$ quiver theory.}}{\Qlb{newheightexample}}%
%{newheightexample.eps}{\special{ language "Scientific Word";  type "GRAPHIC";
%width 2.5396in;  height 1.4687in;  depth 0pt;  original-width 5.9626in;
%original-height 3.4329in;  cropleft "0";  croptop "1";  cropright "1";
%cropbottom "0";  filename 'newheightexample.eps';file-properties "XNPEU";}}}%
%BeginExpansion
%EndExpansion

A dimer model corresponds to a collection of black and white atoms on a
bipartite lattice\footnote{This is a lattice which admits a coloring of its
vertices by black and white atoms so that no white atom connects to a black
atom and vice versa.}. \ The atoms of the lattice correspond to vertices, the
oriented links between black and white atoms are edges, and polygons
constructed from the edges are faces of the dimer model. \ A perfect matching
$PM$ is defined as a collection of edges such that each vertex touches a
single edge of $PM$.

For dimers on the plane, every perfect matching $PM$ defines a unique integer
valued height function $h_{PM}$ up to the addition of a constant. \ Given two
faces $A$ and $B$ with common edge $e$, we have:%
\begin{equation}
h_{PM}(A)-h_{PM}(B)=\left\{
\begin{array}
[c]{c}%
\pm1\text{ if }e\notin PM\\
\mp2\text{ if }e\in PM
\end{array}
\right\}
\end{equation}
where the overall sign is fixed by the orientation of $e$. \ The function
$h_{PM}$ determines the profile of a stepped surface.

\begin{figure}
[ptb]
\begin{center}
\includegraphics[
height=1.9327in,
width=2.0681in
]%
{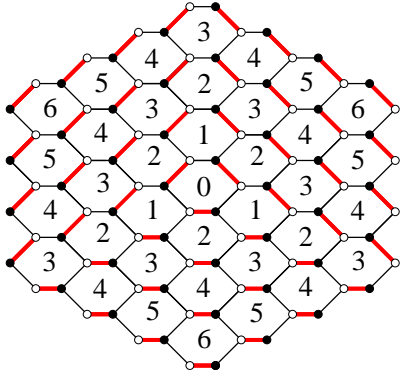}%
\caption{The empty room perfect matching (red). \ The integers at the center
of each face indicate the value of the height function. \ These heights map to
rank assignments in the $\mathbb{C} ^{3}/\mathbb{Z} _{n}\times\mathbb{Z} _{n}$
quiver theory given by the graph dual of the dimer lattice shown in figure
(\ref{graphdualquiver}).}%
\label{emptyrankswithatoms}%
\end{center}
\end{figure}

\begin{figure}
[ptbptb]
\begin{center}
\includegraphics[
height=2.4226in,
width=2.5928in
]%
{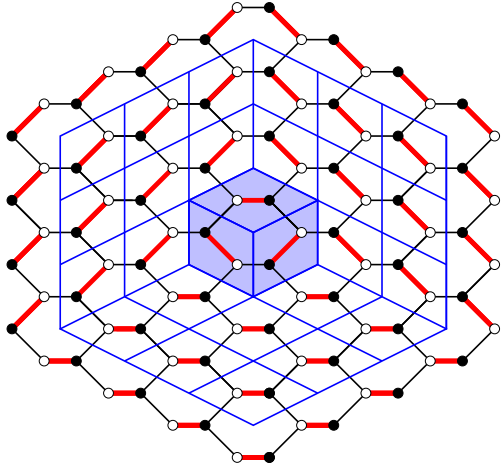}%
\caption{By surrounding each edge belonging to a perfect matching (red) by a
rhombus (blue), we obtain the profile for the corresponding crystal melting
configuration. \ The figure shows the single box crystal melting configuration
(shaded).}%
\label{oneboxcrystal}%
\end{center}
\end{figure}

Three dimensional crystal melting is specified by perfect matchings of the
infinite honeycomb lattice. \ The perfect matching for the frozen crystal with
no atoms removed is shown in figure (\ref{emptyrankswithatoms}).%
%TCIMACRO{\FRAME{ftbpFU}{2.0681in}{1.9327in}{0pt}{\Qcb{The empty room perfect
%matching (red). \ The integers at the center of each face indicate the value
%of the height function. \ These heights map to rank assignments in the
%$\U{2102} ^{3}/\U{2124} _{n}\times\U{2124} _{n}$ quiver theory given by the
%graph dual of the dimer lattice shown in figure (\ref{graphdualquiver}%
%).}}{\Qlb{emptyrankswithatoms}}{emptyrankswithatoms.eps}%
%width 2.0681in;  height 1.9327in;  depth 0pt;  original-width 3.6886in;
%original-height 3.4454in;  cropleft "0";  croptop "1";  cropright "1";
%cropbottom "0";  filename 'emptyrankswithatoms.eps';file-properties "XNPEU";}%
%}}%
%BeginExpansion
%EndExpansion
To visualize the crystal, it is helpful to draw a rhombus around each edge
contained in the perfect matching. \ The corresponding stepped surface
constructed from these rhombi realizes the profile of the melted crystal.
\ See figure (\ref{oneboxcrystal}) for an example.%
%TCIMACRO{\FRAME{ftbpFU}{2.5928in}{2.4226in}{0pt}{\Qcb{By surrounding each edge
%belonging to a perfect matching (red) by a rhombus (blue), we obtain the
%profile for the corresponding crystal melting configuration. \ The figure
%shows the single box crystal melting configuration (shaded).}}%
%{\Qlb{oneboxcrystal}}{oneboxcrystal.eps}%
%width 2.5928in;  height 2.4226in;  depth 0pt;  original-width 3.6886in;
%original-height 3.4454in;  cropleft "0";  croptop "1";  cropright "1";
%cropbottom "0";  filename 'oneboxcrystal.eps';file-properties "XNPEU";}}}%
%BeginExpansion
%EndExpansion%
%TCIMACRO{\FRAME{ftbpFU}{6.0739in}{2.0531in}{0pt}{\Qcb{The graph dual of the
%infinite honeycomb lattice defines a quiver gauge theory. \ The quiver nodes
%determine a lattice in the complex plane with generators $e_{1}=e^{-i\pi/6}$
%and $e_{2}=e^{i\pi/2}$. \ Superpotential terms are given by signed chiral
%gauge invariant operators constructed from triangles enclosing a single $+$ or
%$-$ .}}{\Qlb{graphdualquiver}}{newergraphdualquiver.eps}%
%width 6.0739in;  height 2.0531in;  depth 0pt;  original-width 7.3864in;
%original-height 2.4782in;  cropleft "0";  croptop "1";  cropright "1";
%cropbottom "0";  filename 'newergraphdualquiver.eps';file-properties "XNPEU";}%
%}}%
%BeginExpansion
\begin{figure}
[ptbptbptb]
\begin{center}
\includegraphics[
height=2.0531in,
width=6.0739in
]%
{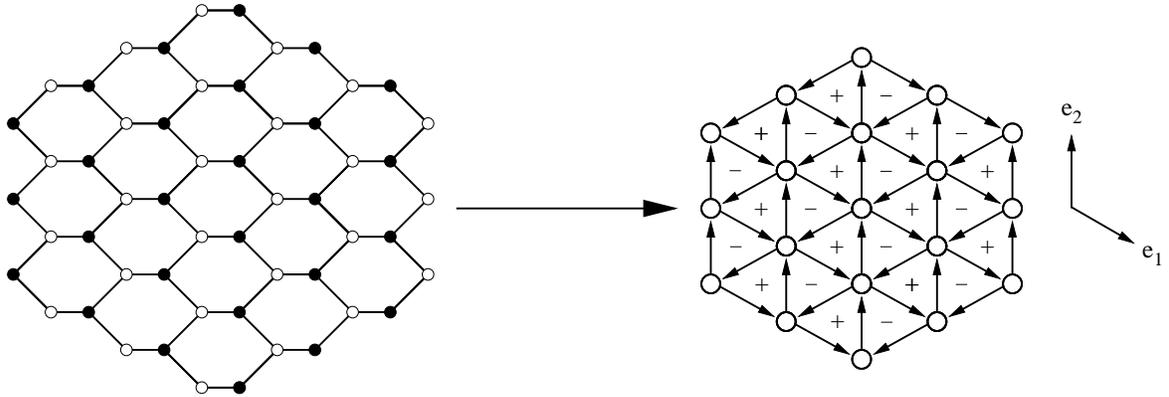}%
\caption{The graph dual of the infinite honeycomb lattice defines a quiver
gauge theory. \ The quiver nodes determine a lattice in the complex plane with
generators $e_{1}=e^{-i\pi/6}$ and $e_{2}=e^{i\pi/2}$. \ Superpotential terms
are given by signed chiral gauge invariant operators constructed from
triangles enclosing a single $+$ or $-$ .}%
\label{graphdualquiver}%
\end{center}
\end{figure}
%EndExpansion

In an unrelated development, it has been shown that dimers on a $T^{2}$
determine the low energy dynamics of type IIB space-time filling D3-branes
probing a toric Calabi-Yau singularity
\cite{HananyKennaway,HananydBranedimers,HananyGaugefromTorictiling}. \ The
physical origin of this $T^{2}$ was recently interpreted using mirror symmetry
in \cite{VafaKennaway}. \ In a gauge theory dimer, the faces determine gauge
groups, the oriented edges shared by faces give the bifundamental matter, and
the vertices give all tree level superpotential terms. \ Note that by
infinitely extending any finite dimer model on $T^{2}$, we obtain a dimer
model which tiles the plane. \ Indeed, the original motivation for this note
was to give a physical interpretation of the infinite honeycomb lattice dimer
model of crystal melting purely as a quiver gauge theory of the type shown in
figure (\ref{graphdualquiver}).

The plan of this note is as follows. \ In section \ref{Twodim} we study flops
of black hole charge configurations in the local geometry $%
%TCIMACRO{\U{2102} }%
%BeginExpansion
\mathbb{C}
%EndExpansion
^{2}/%
%TCIMACRO{\U{2124} }%
%BeginExpansion
\mathbb{Z}
%EndExpansion
_{n}$ which realize 2d crystal melting configurations. \ We next show in
section \ref{3dtab} that at the level of homology in the type IIB\ mirror of $%
%TCIMACRO{\U{2102} }%
%BeginExpansion
\mathbb{C}
%EndExpansion
^{3}/%
%TCIMACRO{\U{2124} }%
%BeginExpansion
\mathbb{Z}
%EndExpansion
_{n}\times%
%TCIMACRO{\U{2124} }%
%BeginExpansion
\mathbb{Z}
%EndExpansion
_{n}$, there are naively far too many candidate charge configurations to match
to 3d crystal melting. \ Using technology from gauge theory dimer models, in
section \ref{fracdimers} we present a refined analysis in the bounded derived
category of coherent sheaves which recovers the expected correspondence.
\ Following this, in section \ref{PartitionFunctions} we connect the partition
functions for crystal melting to the representation theory of algebras
naturally associated to the duality groups of the orbifold theories and give a
physical interpretation of this counting. \ Finally, in section
\ref{Topconnection} we explain the direct mathematical connection between our
ensemble of black hole crystal melting configurations and topological string
theory. \ We then conclude and discuss possible extensions of this work.

\section{2d Crystals and Black Holes \label{Twodim}}

Near the orbifold point, the gauged quiver quantum mechanics of branes
wrapping cycles in $%
%TCIMACRO{\U{2102} }%
%BeginExpansion
\mathbb{C}
%EndExpansion
^{2}/%
%TCIMACRO{\U{2124} }%
%BeginExpansion
\mathbb{Z}
%EndExpansion
_{\infty}\times%
%TCIMACRO{\U{2102} }%
%BeginExpansion
\mathbb{C}
%EndExpansion
$ is given by an infinite one dimensional lattice of $U(N_{i})$ quiver nodes
attached by chiral superfields $X_{i,i+1}$, $Y_{i+1,i}$ and $Z_{i}$
transforming in the representations $(N_{i},\overline{N}_{i+1})$,
$(N_{i+1},\overline{N}_{i})$ and $(N_{i},\overline{N}_{i})$,
respectively.\ \ D-brane probes of the finite $n$ orbifold $%
%TCIMACRO{\U{2102} }%
%BeginExpansion
\mathbb{C}
%EndExpansion
^{2}/%
%TCIMACRO{\U{2124} }%
%BeginExpansion
\mathbb{Z}
%EndExpansion
_{n}$ were first studied in \cite{DouglasMoore}. \ The superpotential of the
theory is:%
\begin{equation}
W=\underset{i\in%
%TCIMACRO{\U{2124} }%
%BeginExpansion
\mathbb{Z}
%EndExpansion
}{\sum}Tr\left(  Z_{i}X_{i,i+1}Y_{i+1,i}-Z_{i+1}Y_{i+1,i}X_{i,i+1}\right)  .
\label{2dsuperpot}%
\end{equation}
This is known as the $A_{\infty}^{\infty}$ quiver \cite{BensonQuivers}. \ The
\textquotedblleft fractional branes\textquotedblright\ of the orbifold CFT
correspond to wrapping D2-branes over the blown up 2-cycles given by the
simple roots $\left\{  \alpha_{i}\right\}  _{i\in%
%TCIMACRO{\U{2124} }%
%BeginExpansion
\mathbb{Z}
%EndExpansion
}$ of the associated A-series Lie algebra. \ The intersection product of these
cycles is up to a minus sign given by the Cartan matrix of the algebra.%
%TCIMACRO{\FRAME{ftbpFU}{4.6284in}{0.5687in}{0pt}{\Qcb{In 2d tableau quivers,
%the vertical height of the 2d tableau determines the ranks of the gauge
%groups. \ In the figure, the ranks decrease in the direction of the red
%arrows.}}{\Qlb{2dtabranksthin}}{2dtabranksthin.eps}%
%width 4.6284in;  height 0.5687in;  depth 0pt;  original-width 8.8136in;
%original-height 1.0585in;  cropleft "0";  croptop "1";  cropright "1";
%cropbottom "0";  filename '2dtabranksthin.eps';file-properties "XNPEU";}}}%
%BeginExpansion
\begin{figure}
[ptb]
\begin{center}
\includegraphics[
height=0.5687in,
width=4.6284in
]%
{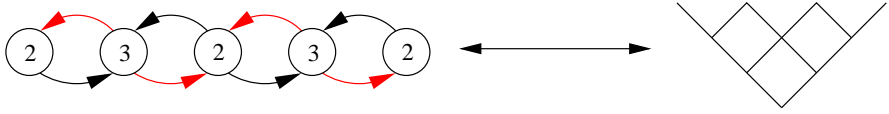}%
\caption{In 2d tableau quivers, the vertical height of the 2d tableau
determines the ranks of the gauge groups. \ In the figure, the ranks decrease
in the direction of the red arrows.}%
\label{2dtabranksthin}%
\end{center}
\end{figure}
%EndExpansion
\ The charge vector:%
\begin{equation}
Q=\underset{i}{\sum}N_{i}\alpha_{i}%
\end{equation}
defines a quiver with gauge group $U(N_{i})$ at the $i^{th}$ node. \ We define
a 2d tableau quiver as one where the ranks correspond to the heights of a 2d
crystal. \ 

The flop of the $\mathbb{P}^{1}$ corresponding to $\alpha_{k}$ is given by the
Weyl reflection:%
\begin{equation}
\sigma_{\alpha_{k}}(Q)=Q-(\alpha_{k}\cdot Q)\alpha_{k}%
\end{equation}
which in four dimensional gauge theories describes a Seiberg-like duality of
the $k^{th}$ quiver node \cite{CFIKV}. \ From the perspective of the quiver
theory, this flop corresponds to the gauge coupling $g_{k}^{-2}$ becoming
negative. \ The ranks in the flopped theory are determined by conservation of
flux, or equivalently by the brane creation mechanism:%
\begin{equation}
N_{i}^{\prime}=N_{i}+\left(  N_{k+1}+N_{k-1}-2N_{k}\right)  \delta_{i,k}.
\label{Seibdual}%
\end{equation}
The duality group of the quiver theory is generated by products of the
$\sigma_{\alpha_{i}}$ and is isomorphic to the Weyl group of the infinite
A-series algebra.

Our goal is to show that the flops of the empty room brane configuration:%
\begin{equation}
Q_{empty}=\underset{i}{\sum}\left\vert i\right\vert \alpha_{i}%
\end{equation}
correspond to 2d crystal melting configurations. \ Weyl reflection by
$\sigma_{\alpha_{k}}$ yields:%
\begin{equation}
\sigma_{\alpha_{k}}\left(  Q_{empty}\right)  =Q_{empty}+2\delta_{k,0}%
\alpha_{0}. \label{rankle}%
\end{equation}
But this is precisely the rule for adding a box to the crystal melting
configuration. \ Indeed, the process of adding a box to the tableau changes
the height of the $0^{th}$ quiver node from $0$ to $2$. \ Note that the
particular choice of ranks made in the empty room configuration effectively
constrains the location of the non-trivial flops. \ 

We now show that this correspondence between crystal melting and flops of 2d
tableau quivers holds more generally. \ Given a quiver node with rank $M$,
there are four possible rank assignments consistent with equation
(\ref{derivative}):%
\begin{equation}
(M-1,M,M+1),(M+1,M,M-1),(M+1,M,M+1),(M-1,M,M-1). \label{last}%
\end{equation}
It follows from equation (\ref{Seibdual}) that in all cases a 2d tableau
quiver always flops to another 2d tableau quiver. \ Furthermore, given two 2d
tableau quivers $T$ and $T^{\prime}=T+2\alpha_{k}$:%
\begin{equation}
\sigma_{\alpha_{k}}\left(  T\right)  =T+2\alpha_{k}=T^{\prime}.
\end{equation}
This implies that by acting with the duality group on the empty room
configuration we can reach any 2d tableau quiver. \ A similar set of
statements holds for the inverted empty room quiver:%
\begin{equation}
Q_{inv}=\underset{i}{\sum}\left(  N-\left\vert i\right\vert \right)
\alpha_{i} \label{invertedTab}%
\end{equation}
where $N\rightarrow\infty$ faster than $n\rightarrow\infty$ in $%
%TCIMACRO{\U{2102} }%
%BeginExpansion
\mathbb{C}
%EndExpansion
^{2}/%
%TCIMACRO{\U{2124} }%
%BeginExpansion
\mathbb{Z}
%EndExpansion
_{n}$.

Finally, in the case where $n$ is large but finite, the quiver is given by a
periodic one dimensional lattice. \ The empty room black hole charge
configuration now has two corners about which crystal melting takes place.
\ At finite $n$, the corresponding 2d tableaux begin to mix after sufficiently
many flops. \ In \cite{BabyUniverse} a similar mixing of 2d tableaux was
interpreted as the non-perturbative creation of baby universes with
Bertotti-Robinson cosmology $AdS_{2}\times S^{2}$.

\subsection{Duality Cascades}

This same quiver also defines an $\mathcal{N}=2$ four dimensional gauge theory
described by type IIB\ D5-branes wrapped over the collapsed 2-cycles of the
geometry. \ The flop transitions considered above correspond to Seiberg-like
dualities of the quiver gauge theory. \ The 1-loop exact $i^{th}$ holomorphic
gauge coupling at scale $\mu$ is:%
\begin{equation}
\frac{1}{g_{i}^{2}\left(  \mu\right)  }=\frac{(2N_{i}-N_{i+1}-N_{i-1})}%
{8\pi^{2}}\log\left(  \frac{\mu}{\Lambda_{i}}\right)  .
\end{equation}
This implies that local maxima in the ranks of the quiver gauge theory are
asymptotically free. \ 

Adding mass terms of the form $TrZ_{i}^{2}$ for all $i$ to the superpotential
of equation (\ref{2dsuperpot}) breaks the system to an $\mathcal{N}=1$ gauge
theory of the type studied in \cite{CFIKV}. \ Starting from the $Q_{inv}$ rank
assignments of equation (\ref{invertedTab}), the theory will flow into the
infrared, at which point it becomes appropriate to Seiberg dualize the
corresponding gauge group. \ In this way, crystal melting corresponds to a
duality cascade in renormalization group flow. \ Although we omit the details,
it is also possible to show that by a suitable choice of couplings in the
initial configuration, any sequence of adding boxes to an inverted 2d tableau
can be achieved. \ In contrast to the original case of duality cascades on the
conifold considered in \cite{KlebanovStrassler}, the cascade of this system
can continue indefinitely before it confines.

\section{3d Crystals and Black Holes \label{3dtab}}

We now consider black hole charges parametrized by three dimensional crystal
melting configurations. \ Interpreting the infinite honeycomb lattice as a
gauge theory dimer model of the type described in the introduction, the
procedure of \cite{DouglasMoore} shows that the associated gauged quiver
quantum mechanics system probes the supersymmetric orbifold $%
%TCIMACRO{\U{2102} }%
%BeginExpansion
\mathbb{C}
%EndExpansion
^{3}/%
%TCIMACRO{\U{2124} }%
%BeginExpansion
\mathbb{Z}
%EndExpansion
_{n}\times%
%TCIMACRO{\U{2124} }%
%BeginExpansion
\mathbb{Z}
%EndExpansion
_{n}$ in the limit of large $n$. \ The quiver nodes determine a lattice in the
complex plane with generators $e_{1}=e^{-i\pi/6}$ and $e_{2}=e^{i\pi/2}$ and
correspond to a basis of branes wrapping collapsed 2- and 4-cycles. \ For
faces $F=ae_{1}+be_{2}$ with $a,b\in%
%TCIMACRO{\U{2124} }%
%BeginExpansion
\mathbb{Z}
%EndExpansion
$, the matter content is given by bifundamentals $X_{F,F+e_{1}},$
$Y_{F,F+e_{2}}$ and $Z_{F,F-e_{1}-e_{2}}$ in the obvious notation. \ The
superpotential is:%
\begin{equation}
W=\underset{F}{\sum}Tr\left(  X_{F,F+e_{1}}Y_{F+e_{1},F+e_{1}+e_{2}}%
Z_{F+e_{1}+e_{2},F}-X_{F,F+e_{1}}Z_{F+e_{1},F-e_{2}}Y_{F-e_{2},F}\right)  .
\end{equation}
We define a 3d tableau quiver as one where the rank assignments are given by
the height function of the dimer model. \ See figure (\ref{graphdualquiver})
for a picture of this quiver theory.

For $%
%TCIMACRO{\U{2102} }%
%BeginExpansion
\mathbb{C}
%EndExpansion
^{3}/%
%TCIMACRO{\U{2124} }%
%BeginExpansion
\mathbb{Z}
%EndExpansion
_{n}\times%
%TCIMACRO{\U{2124} }%
%BeginExpansion
\mathbb{Z}
%EndExpansion
_{n}$, almost all of the flops of the geometry will alter the topology of the
quiver and thus the classical geometry probed by the branes. \ In this sense,
the duality group of the 2d tableau quivers is quite special because each flop
preserves the intersection product of the cycles in the geometry. \ Even with
this restriction, a preliminary analysis reveals far too many duals of the
empty room quiver in comparison with crystal melting.

Because the analysis to follow is somewhat lengthy, we now give an outline of
each component of sections \ref{3dtab} and \ref{fracdimers}. \ In subsection
\ref{hom} we present a naive analysis of geometric transitions which preserve
the Calabi-Yau geometry and find far too many black hole charge configurations
in comparison with crystal melting configurations.\ \ In section
\ref{fracdimers} we rectify this by determining the physical meaning of the
internal perfect matchings in a general gauge theory dimer model. \ As
explained in subsection \ref{physfrac}, a candidate collection of fractional
branes may contain ghost fields which render the theory unphysical. \ In
subsection \ref{PMexcep} we show that the perfect matchings of the dimer model
parametrize the physical quiver theories. \ In appendix B we compute the
effect of switching from one perfect matching to another and find that it
corresponds to a subset of the transformations considered in subsection
\ref{hom}. \ Returning to the problem of crystal melting, we show in
subsection \ref{crystrev} that the BPS\ black holes generated by these
transformations are in one to one correspondence with crystal melting
configurations. \ To complete our analysis, in subsection \ref{extras} we
discuss the physical meaning of the extra charge configurations obtained in
subsection \ref{hom}.

\subsection{Homology Cycles in the Mirror\label{hom}}

To realize the flop transitions geometrically we pass to the type IIB\ theory
on the mirror manifold $\widetilde{X}$ where worldsheet instanton corrections
have been resummed into the geometry. \ The D0-, D2- and D4-branes which base
the quiver theory in $%
%TCIMACRO{\U{2102} }%
%BeginExpansion
\mathbb{C}
%EndExpansion
^{3}/%
%TCIMACRO{\U{2124} }%
%BeginExpansion
\mathbb{Z}
%EndExpansion
_{n}\times%
%TCIMACRO{\U{2124} }%
%BeginExpansion
\mathbb{Z}
%EndExpansion
_{n}$ map to D3-branes wrapping homology 3-spheres $\Delta_{i}\in
H_{3}(\widetilde{X},%
%TCIMACRO{\U{2124} }%
%BeginExpansion
\mathbb{Z}
%EndExpansion
)$ for $i=1,...,n^{2}$. \ The anti-symmetric intersection pairing $\Delta
_{i}\cap\Delta_{j}$ determines the number of bifundamentals shared by
$\Delta_{i}$ and $\Delta_{j}$. \ The charge vector:%
\begin{equation}
Q=\underset{i}{\sum}N_{i}\Delta_{i}%
\end{equation}
defines a quiver with gauge group $U(N_{i})$ at the $i^{th}$ node. \ In the
context of Seiberg dualities of four dimensional gauge theories, flops in the
mirror theory were studied in \cite{CFIKV,HananyIqbalMirror}. \ We flop the
brane wrapping $\Delta_{1}$ by passing it through all incoming branes
($\Delta_{1}\cap\Delta_{in}<0$) for $S_{L}$ duality and all outgoing branes
($\Delta_{1}\cap\Delta_{out}>0$) for $S_{R}$ duality.

In the case of the infinite honeycomb lattice, we label the outgoing branes as
$\Delta_{2},\Delta_{3},\Delta_{4}$ and the incoming branes as $\Delta
_{5},\Delta_{6},\Delta_{7}$. \ Applying the transformation $S_{R}$ yields:%
\begin{align}
S_{R}\left(  \Delta_{1}\right)   &  =-\Delta_{1}\\
S_{R}\left(  \Delta_{out}\right)   &  =\Delta_{out}+\left(  \Delta_{1}%
\cap\Delta_{out}\right)  \Delta_{1}=\Delta_{out}+\Delta_{1}%
\end{align}
where all the other $\Delta_{i}$ which base the quiver remain unchanged.
\ This new basis of branes does not preserve the intersection product of the
geometry. \ We perform a further $S_{R}$ transformation by passing the brane
wrapping $-\Delta_{1}$ through $\Delta_{5},\Delta_{6},\Delta_{7}$ to find:%
\begin{align}
S_{R}^{2}\left(  \Delta_{1}\right)   &  =-\left(  -\Delta_{1}\right)
=\Delta_{1}\\
S_{R}^{2}\left(  \Delta_{out}\right)   &  =\Delta_{out}+\Delta_{1}\\
S_{R}^{2}\left(  \Delta_{in}\right)   &  =\Delta_{in}+\left(  -\Delta_{1}%
\cap\Delta_{in}\right)  \left(  -\Delta_{1}\right)  =\Delta_{in}-\Delta_{1}%
\end{align}
or,%
\begin{equation}
S_{R}^{2}\left(  \Delta_{j}\right)  =\Delta_{j}+\left(  \Delta_{1}\cap
\Delta_{j}\right)  \Delta_{1} \label{PLeffect}%
\end{equation}
for all $j$. \ The intersection product is now unchanged:%
\begin{equation}
S_{R}^{2}\left(  \Delta_{i}\right)  \cap S_{R}^{2}\left(  \Delta_{j}\right)
=\Delta_{i}\cap\Delta_{j}.
\end{equation}
Conservation of flux determines the ranks of the gauge groups:%
\begin{equation}
\underset{i=1}{\overset{7}{\sum}}N_{i}\Delta_{i}=S_{R}^{2}\left(
N_{1}\right)  \Delta_{1}+\underset{i=2}{\overset{4}{\sum}}S_{R}^{2}\left(
N_{i}\right)  \left(  \Delta_{i}+\Delta_{1}\right)  +\underset{i=5}%
{\overset{7}{\sum}}S_{R}^{2}\left(  N_{i}\right)  \left(  \Delta_{i}%
-\Delta_{1}\right)
\end{equation}
so that:%
\begin{equation}
S_{R}^{2}\left(  N_{i}\right)  =N_{i}+\delta_{i,1}\left(  \sum N_{in}-\sum
N_{out}\right)  \label{rankchanger}%
\end{equation}
in the obvious notation. \ A similar analysis yields:%
\begin{equation}
S_{L}^{2}\left(  N_{i}\right)  =N_{i}+\delta_{i,1}\left(  \sum N_{out}-\sum
N_{in}\right)  .
\end{equation}

Now apply the $S_{R}^{2}$ transformation to any node of the empty room quiver
with rank assignments as in figure (\ref{emptyrankswithatoms}). \ First
consider any face in the dimer model with two edges in the perfect matching.
\ In this case, inspection of figure (\ref{emptyrankswithatoms}) implies that
$S_{R}^{2}(N_{1})=N_{1}$. \ For the face with three edges in the perfect
matching, $N_{1}=0$, $N_{2,3,4}=1$ and $N_{5,6,7}=2$ so that:%
\begin{equation}
S_{R}^{2}\left(  N_{1}\right)  =0+(2+2+2-1-1-1)=+3\text{.}%
\end{equation}
This matches the change in the height function from crystal melting. \ Note
that the transformation $S_{L}^{2}$ would have produced a negative rank gauge group.

To show that any crystal melting black hole charge configuration may be
reached by a sequence of $S_{R}^{2}$ transformations, dualize the face of a 3d
tableau quiver where all incoming edges belong to the perfect matching. \ The
rank assignments are:%
\begin{align}
\ N_{F+e_{i}}  &  =N_{F-e_{1}-e_{2}}=N_{F}+1\text{ (outgoing)}\\
\ N_{F-e_{i}}  &  =N_{F+e_{1}+e_{2}}=N_{F}+2\text{ (incoming)}%
\end{align}
for $i=1,2$. \ Applying the transformation $S_{R}^{2}$ yields:%
\begin{equation}
\ S_{R}^{2}\left(  N_{F}\right)  =N_{F}+\left(  \sum N_{in}-\sum
N_{out}\right)  =N_{F}+3\text{.} \label{rankaddabox}%
\end{equation}
This matches the change in height of the crystal. \ By induction, we can reach
any 3d tableau quiver by successive $S_{R}^{2}$ transformations of the empty
room quiver.

But at the level of homology, this is not the full collection of rank
assignments which $S_{R}^{2}$ generates. \ Applying the transformation
$S_{R}^{2}$ $n$ times yields:%
\begin{equation}
\ \left(  S_{R}^{2}\right)  ^{n}\left(  N_{F}\right)  =N_{F}+3n
\end{equation}
which is a much larger spectrum of BPS\ black holes. \ As we show in section
\ref{fracdimers}, \textit{none} of these extraneous charge configurations are stable.

\section{Fractional Branes and Dimers \label{fracdimers}}

To obtain a more refined physical description of admissible $S_{R}^{2}$
transformed brane configurations, we pass to the bounded derived category of
coherent sheaves on $X=%
%TCIMACRO{\U{2102} }%
%BeginExpansion
\mathbb{C}
%EndExpansion
^{3}/%
%TCIMACRO{\U{2124} }%
%BeginExpansion
\mathbb{Z}
%EndExpansion
_{n}\times%
%TCIMACRO{\U{2124} }%
%BeginExpansion
\mathbb{Z}
%EndExpansion
_{n},$ denoted $D^{b}(X)$. \ Under certain plausible assumptions, most of the
results of the following subsections hold for general toric Calabi-Yau
threefolds. \ Although a candidate collection of fractional branes in
$D^{b}(X)$ may generate the correct quiver structure, the corresponding gauge
theory must not contain any ghosts. \ Indeed, in subsection \ref{hom} we
merely counted the number of bifundamentals given by the intersection product
of homology cycles in the mirror and did not address the physical properties
of these massless degrees of freedom. \ Our goal in this section is to
eliminate all unphysical candidate collections.

\begin{figure}
[ptb]
\begin{center}
\includegraphics[
height=1.4014in,
width=4.1552in
]%
{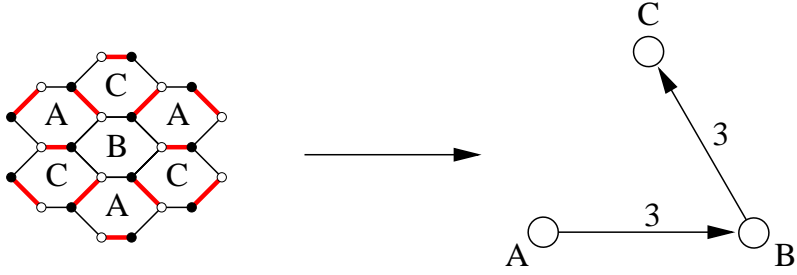}%
\caption{The gauge theory dimer model for the supersymmetric orbifold
$\mathbb{C} ^{3}/\mathbb{Z} _{3}$ is shown on the left. \ By deleting the
edges belonging to the internal perfect matching (red) we obtain a Beilinson
quiver which by definition contains no directed loops.}%
\label{newz3orbifold}%
\end{center}
\end{figure}

A physical basis of fractional branes is determined by a collection of
exceptional sheaves supported on a complex surface $V$ obtained from a partial
resolution of $X$. \ An exceptional collection may be thought of as a basis of
branes wrapping $V$ with the property that the associated quiver theory is
obtained from the physical quiver by deleting a minimal number of
bifundamentals so that no directed loops remain. \ An exceptional collection
is called strong when the basis of fractional branes it generates contains no
ghost matter. \ In this case, the quiver with deleted arrows is called a
Beilinson quiver. \ See figure (\ref{newz3orbifold}) for an example. \ Note
that a Beilinson quiver has a certain number of starting (terminal) quiver
nodes such that all attached bifundamentals are outgoing (incoming). \ We
review in appendix A the precise definition of strong exceptional collections
and their relation to fractional branes.%
%TCIMACRO{\FRAME{ftbpFU}{4.1552in}{1.4014in}{0pt}{\Qcb{The gauge theory dimer
%model for the supersymmetric orbifold $\U{2102} ^{3}/\U{2124} _{3}$ is shown
%on the left. \ By deleting the edges belonging to the internal perfect
%matching (red) we obtain a Beilinson quiver which by definition contains no
%directed loops.}}{\Qlb{newz3orbifold}}{newz3orbifold.eps}%
%width 4.1552in;  height 1.4014in;  depth 0pt;  original-width 6.3296in;
%original-height 2.1179in;  cropleft "0";  croptop "1";  cropright "1";
%cropbottom "0";  filename 'newZ3orbifold.eps';file-properties "XNPEU";}}}%
%BeginExpansion
%EndExpansion

Although implementing these extra physical conditions appears unwieldy, it has
recently been shown that in the case of gauge theory dimer models, the perfect
matchings of the dimer model parametrize these exceptional collections
\cite{HHVExceptional}. \ As we show in appendix B, the local rearrangement of
a perfect matching corresponds to the action of $S_{R}^{2}$ or $S_{L}^{2}$ on
the fractional branes. \ This allows us to recover the expected match to
crystal melting. \ Based on the general analysis presented in sections
\ref{physfrac} and \ref{PMexcep}, it follows that the additional $S_{R,L}^{2}$
transformations of section \ref{hom} introduce ghost matter into the quiver.
\ We conjecture that the corresponding bound state decays before these ghosts
are produced and explain how this picture is consistent with expectations from supergravity.

\subsection{Physical Fractional Branes and Ghosts \label{physfrac}}

The \textquotedblleft fractional branes\textquotedblright\ of a physical
quiver theory correspond to branes wrapping collapsed cycles in the geometry.
\ The matter content of the quiver theory is determined by massless open
strings with the appropriate Chan-Paton factors. \ More formally, open strings
in the quiver are described by maps of the form Hom$_{D^{b}(X)}^{k}\left(
E,F\right)  $ for $k=0,1,2,3$ with $E$ and $F$ in $D^{b}(X)$
\cite{DouglasCategories}. \ At the orbifold point of the theory, the mass
squared of the corresponding zero modes is $M_{k}^{2}=\left(  k-1\right)  /2$
in string units. \ Although the GSO\ projection removes the $k=0$ (and $k=2$)
zero modes from the string spectrum, there are still potential instabilities
in the brane system. \ A general collection of candidate fractional branes in
$D^{b}(X)$ may also contain more exotic massless degrees of freedom given by
bifundamental vector bosons corresponding to the first excited state of the
map Hom$_{D^{b}(X)}^{0}\left(  E,F\right)  $. \ These fields were interpreted
in \cite{VafaSupergroup} as tachyons in a topological brane anti-brane system.
\ In the physical string theory, such fields signal the presence of uncanceled
ghosts \cite{WijnholtGhosts,GhostBranes}.

Whereas gauge invariance and the Ward identities ensure that all negative norm
states of an adjoint valued vector boson decouple from the physical Hilbert
space, no such decoupling occurs when the vector boson is in the bifundamental
of two gauge groups. \ Indeed, the ghost number of the corresponding vertex
operators in the open topological B-model shows that such fields and thus the
associated quiver theories are unphysical \cite{WijnholtGhosts}. \ In the case
of gauged quiver quantum mechanics, these exotic bosonic propagating degrees
of freedom correspond to bifundamental\footnote{For these more exotic
bifundamentals, the orientation of the arrow in the quiver theory is aligned
in the opposite direction to that of the associated Hom$^{0}$ map
\cite{WijnholtGhosts}.} scalars with a wrong sign kinetic term.

The kinetic term of either a physical or unphysical bifundamental scalar $X$
is:%
\begin{equation}
\ L_{kin}=c\left(  \psi\right)  Tr\left(  \left(  D_{0}X\right)  \left(
D_{0}X\right)  ^{\dag}\right)  \label{ghostkin}%
\end{equation}
where the real function $c\left(  \psi\right)  $ is a moduli dependent factor
which is positive for physical matter and negative for ghost matter. \ The
monodromy transformations $S_{R,L}^{2}$ will therefore alter $c\left(
\psi\right)  $ and may even cause it to switch sign. \ Before this occurs, the
system enters a regime of strong coupling where $c\left(  \psi\right)  $ is
nearly zero. \ Although we do not understand the dynamics at small $c\left(
\psi\right)  $, we conjecture that the corresponding bound state decays before
the unphysical ghost is produced. \ We will return to this issue in subsection
\ref{extras}.

\subsection{Perfect Matchings and Exceptional Collections \label{PMexcep}}

To construct a basis of fractional branes which excludes the presence of ghost
matter, it suffices to consider strong exceptional collections of sheaves
supported on a complex surface $V$ which may be used to generate a physical
basis of fractional branes, as in \cite{CFIKV,WijnholtLargeVol}. \ We now
explain the connection between perfect matchings and exceptional collections.

To partially classify the perfect matchings of the dimer model, fix a
reference perfect matching $PM_{0}$ and consider the formal difference of
edges given by $PM-PM_{0}$ where $PM$ is any other perfect matching. \ This
formal difference determines a homology 1-cycle in the $T^{2}$ of the dimer
model. \ As shown in \cite{KOSDimerReview}, the corresponding subset of
lattice points in $H_{1}\left(  T^{2},%
%TCIMACRO{\U{2124} }%
%BeginExpansion
\mathbb{Z}
%EndExpansion
\right)  \simeq%
%TCIMACRO{\U{2124} }%
%BeginExpansion
\mathbb{Z}
%EndExpansion
\times%
%TCIMACRO{\U{2124} }%
%BeginExpansion
\mathbb{Z}
%EndExpansion
$ defines a convex polytope. \ We define an internal (external) perfect
matching as one where its lattice point lies in the interior (boundary) of
this polytope. \ The key insight of \cite{HHVExceptional} was that the
internal perfect matchings parametrize all possible ways of deleting \ a
minimal number of arrows in the quiver so that no directed loops remain.
\ Note that this is a necessary condition for forming a Beilinson quiver.

As shown in \cite{HHVExceptional}, up to tensoring all sheaves by a common
line bundle, the internal perfect matchings of the dimer model are in one to
one correspondence with exceptional collections of sheaves which generate the
\textit{same} four dimensional quiver gauge theory. \ On an example by example
basis it was shown in \cite{HHVExceptional} that each such collection is
strong, so we shall assume that each exceptional collection generates a
physical quiver gauge theory\footnote{In fact, achieving the match with
crystal melting configurations does not require this stronger assumption. \ It
follows from the discussion in section \ref{Topconnection} that the empty room
perfect matching defines a strong exceptional collection because all of the
sheaves in this collection are generated by their global sections. \ Because
all of the other internal perfect matchings of crystal melting define
foundations for the same strong helix, we conclude that all of the perfect
matchings of crystal melting generate physical quiver gauge theories for the
large $n$ orbifold $%
%TCIMACRO{\U{2102} }%
%BeginExpansion
\mathbb{C}
%EndExpansion
^{3}/%
%TCIMACRO{\U{2124} }%
%BeginExpansion
\mathbb{Z}
%EndExpansion
_{n}\times%
%TCIMACRO{\U{2124} }%
%BeginExpansion
\mathbb{Z}
%EndExpansion
_{n}$.}. \ In other words: \textit{The internal perfect matchings parametrize
all physical collections of fractional branes which preserve both the
adjacency of bifundamentals in the quiver as well as the form of the
superpotential.}

Now that we have a catalogue of admissible collections of fractional branes in
terms of perfect matchings, we can consider the effect of switching perfect
matchings. \ We defer this computation to appendix B where we show that
changing a starting (resp. terminal) node of the associated Beilinson quiver
to a terminal (resp. starting) node corresponds to a $S_{R}^{2}$ (resp.
$S_{L}^{2}$) transformation.

\subsection{Crystal Melting Revisited \label{crystrev}}

We now specify the empty room quiver by two conditions. \ First, we take the
ranks $N_{F}$ equal to the heights $h_{F}$ of the empty room perfect matching
shown in figure (\ref{emptyrankswithatoms}). \ Second, we stipulate that the
collection of fractional branes used to base the empty room quiver theory be
derived from the same empty room perfect matching. \ This second condition
guarantees that the corresponding Beilinson quiver has exactly one starting
node at $(0,0)$ with smallest rank in the quiver gauge theory.

Because there is a single starting node in the Beilinson quiver, we may only
apply the transformation $S_{R}^{2}$ at the node $(0,0)$. \ This
transformation changes the node $(0,0)$ into a terminal node. \ Further, the
three nodes $(1,0),$ $(0,1)$ and $(-1,-1)$ now correspond to starting nodes in
the transformed Beilinson quiver. \ The change in rank of the $(0,0)$ node
under $S_{R}^{2}$ is given by equation (\ref{rankaddabox}):%
\begin{equation}
\ N_{(0,0)}\mapsto N_{(0,0)}+(N_{in}-N_{out})=N_{(0,0)}+(6-3)=N_{(0,0)}+3.
\end{equation}
We thus see that the change in ranks exactly coincides with the change in
heights from switching perfect matchings. \ Iterating again, there are now
three candidate starting nodes for the Beilinson quiver. \ We may apply
$S_{R}^{2}$ to any of these nodes to obtain another 3d tableau black hole.
\ Combining this with the analysis done around equation (\ref{rankaddabox}),
we conclude that: \textit{The spectrum of BPS\ black hole charge
configurations which are generated by geometry preserving flops of the empty
room charge configuration are parametrized by 3d crystal melting
configurations}.

\subsection{Extra Charges, Attractors and Ghosts \label{extras}}

At a formal level, we have parametrized the collection of fractional brane
configurations which can be used to base the quiver theory. \ Even so, it is
important to identify the explicit physical mechanism which prevents
additional charges from appearing in the single particle BPS\ spectrum. \ To
this end, we review the decay of single particle objects first in the context
of Seiberg-Witten theory and then in the context of the attractor mechanism
for four dimensional Calabi-Yau black holes. \ We conjecture that before the
transformation $S_{R}^{2}$ produces ghost matter, a similar process causes the
brane configuration to decay.

First consider the four dimensional $\mathcal{N}=2$ $SU(2)$ gauge theory
studied by Seiberg and Witten in \cite{SeibergWitten}. \ The moduli space of
the theory is parametrized by the coordinate $u=\left\langle Tr\phi
^{2}\right\rangle $ where $\phi$ denotes the adjoint valued Higgs field. \ The
electric and magnetic charge numbers $(n_{m},n_{e})$ determine the central
charge vector $Z=n_{m}a_{D}+n_{e}a$ where $a(u)$ denotes the scalar component
of the $\mathcal{N}=2$ $U(1)$ \textquotedblleft photon\textquotedblright%
\ vector multiplet and $a_{D}(u)$ is its conjugate magnetic dual. \ \ By a
suitable choice of renormalization scheme, the magnetic monopole with charge
$(1,0)$ becomes massless near the point $u=1$. \ At weak coupling, applying
monodromy transformations about the point $u=1$ produces a tower of
BPS\ charges given by $(1,n)$ and $(-1,-n)$ for integers $n$. \ At strong
coupling, however, only the charge configurations $(1,0)$ and $(1,1)$ remain
in the single particle BPS\ spectrum. \ Indeed, in the process of performing
the monodromy transformation about the point $u=1$, a given BPS state will
cross a curve of marginal stability with $a_{D}/a$ real. \ At this curve, such
a state may decay into a multi-particle state of the form $S_{1}+S_{2}$.
\ This same conclusion was reached in the derived category by more rigorous
means using notions of $\Pi$-stability in \cite{AspinwallKarp}.

This entire discussion embeds as the low energy limit of type II string theory
compactified on a rigid Calabi-Yau threefold. \ From the perspective of four
dimensional Calabi-Yau black holes in $\mathcal{N}=2$ supergravity, the above
truncation on the spectrum is also expected. \ To understand this, we first
recall some facts about the attractor mechanism for four dimensional
BPS\ black holes in asymptotically flat space
\cite{AttractorMech,AttStrom,AttFerrKallone,AttFerrKallTwo}. \ Although the
entropy of such black holes depends on the near horizon values of the vector
multiplet moduli, these values are fixed by the charges of the black hole.
\ The position dependence of the vector multiplet moduli is specified by an
attractor flow in moduli space. \ As explained in
\cite{DenefSplitFlow,DenefGreeneSplitflow}, if an attractor flow defined by a
homology 3-cycle $Q$ passes a branch cut in moduli space produced by a
conifold point where a 3-cycle $\Delta$ shrinks to zero size, it will instead
flow to a fixed point with charge given by Picard Lefschetz theory:%
\begin{align}
Q^{\prime}  &  =Q\pm\left(  \Delta\cap Q\right)  \Delta\\
&  =S_{R,L}^{2}(Q)
\end{align}
where in the above we have used the mirror type IIB\ language.\ \ Now consider
an attractor flow which circles the conifold point $n$ times and naively
produces infinitely many different charge configurations. \ As explained in
\cite{DenefSplitFlow}, after the flow crosses the branch cut for the first
time, a new wall of marginal stability appears which extends out from the
conifold point. \ Upon crossing this wall, the attractor flow will split into
two constituent products, one of which becomes massless as it flows to the
conifold point. \ We thus obtain from a different perspective the same
truncation. \ An analysis of monodromies and $\Pi$-stability in the derived
category was presented in \cite{AspinwallHorjaKarp}.

We conjecture that a similar set of decays occur for 3d tableau BPS\ black
holes. \ Although a full analysis is beyond our reach, we can still sketch an
argument of what we expect to happen. \ Because the action of $S_{R,L}^{2}$
corresponds to a monodromy transformation around a singular point in moduli
space, the coefficient $c(\psi)$ of equation (\ref{ghostkin}) is sensitive to
this change. \ We suspect that this variation in the moduli causes the gauged
quiver quantum mechanics to become strongly coupled before $c(\psi)$ changes
sign. \ To prevent the appearance of ghosts, the brane system must develop a
tachyonic mode and decay to some constituent products.

\section{Black Hole Ensembles and Characters \label{PartitionFunctions}}

As mentioned in the introduction, the untwisted RR D0-brane charge of a black
hole charge configuration realized by branes wrapped on cycles in the
orbifolds $%
%TCIMACRO{\U{2102} }%
%BeginExpansion
\mathbb{C}
%EndExpansion
^{2}/%
%TCIMACRO{\U{2124} }%
%BeginExpansion
\mathbb{Z}
%EndExpansion
_{n}$ and $%
%TCIMACRO{\U{2102} }%
%BeginExpansion
\mathbb{C}
%EndExpansion
^{3}/%
%TCIMACRO{\U{2124} }%
%BeginExpansion
\mathbb{Z}
%EndExpansion
_{n}\times%
%TCIMACRO{\U{2124} }%
%BeginExpansion
\mathbb{Z}
%EndExpansion
_{n}$ is given by the sum over ranks in the gauged quiver quantum mechanics:%
\begin{equation}
Q_{RR}^{(0)}\left(  BH\right)  =\frac{1}{\left\vert \Gamma\right\vert
}\underset{i}{\sum}N_{i}\equiv\frac{d}{\left\vert \Gamma\right\vert
}\varepsilon\left(  BH\right)
\end{equation}
where $\left\vert \Gamma\right\vert =n,n^{2}$ and $d=2,3$ for $%
%TCIMACRO{\U{2102} }%
%BeginExpansion
\mathbb{C}
%EndExpansion
^{2}/%
%TCIMACRO{\U{2124} }%
%BeginExpansion
\mathbb{Z}
%EndExpansion
_{n}$ and $%
%TCIMACRO{\U{2102} }%
%BeginExpansion
\mathbb{C}
%EndExpansion
^{3}/%
%TCIMACRO{\U{2124} }%
%BeginExpansion
\mathbb{Z}
%EndExpansion
_{n}\times%
%TCIMACRO{\U{2124} }%
%BeginExpansion
\mathbb{Z}
%EndExpansion
_{n}$, respectively. \ Summing over all black holes which are generated by
geometry preserving flops of the empty room charge configurations yields the
crystal melting partition functions:%
\begin{equation}
Z_{BPS}=\underset{BH}{\sum}q^{\varepsilon\left(  BH\right)  }=Z_{crystal}%
=\left\{
\begin{array}
[c]{c}%
\underset{n\geq1}{%
%TCIMACRO{\dprod }%
%BeginExpansion
{\displaystyle\prod}
%EndExpansion
}\left(  1-q^{n}\right)  ^{-1}\text{ (2d)}\\
\underset{n\geq1}{%
%TCIMACRO{\dprod }%
%BeginExpansion
{\displaystyle\prod}
%EndExpansion
}\left(  1-q^{n}\right)  ^{-n}\text{ (3d)}%
\end{array}
\right\}  . \label{quiverpartition}%
\end{equation}

These partition functions are characters of the basic representations of the
infinite dimensional algebra $sl_{\infty}$ and the affine algebra
$\widehat{sl}\left(  \infty\right)  $ at large central charge. \ We refer to
appendices C and D for the definitions and representation theory of these
algebras. \ Under the action of the Weyl group, the orbit of the highest
weight $\lambda$ of the basic representation $L\left(  \lambda\right)  $ of an
infinite dimensional algebra $\mathcal{A}$ generates the weight system
$\Omega_{\lambda}$\footnote{This follows from the definition of the Weyl group
of an algebra. \ Essentially quoting from \cite{KACBOMBAY}, the Weyl group $W$
of an algebra $\mathcal{A}$ is the group of automorphisms of the Cartan
subalgebra which are restrictions of conjugations by elements of $A$, the
group obtained via exponentiation of $\mathcal{A}$. \ }. \ In the cases we
study, the multiplicity of each weight in $L\left(  \lambda\right)  $ is
unity. \ It therefore follows that the orbit of $\lambda$ under the action of
the Weyl group generates the representation. \ Because the Weyl groups of the
algebras $sl_{\infty}$ and $\widehat{sl}\left(  \infty\right)  $ coincide with
the duality groups of the $%
%TCIMACRO{\U{2102} }%
%BeginExpansion
\mathbb{C}
%EndExpansion
^{2}/%
%TCIMACRO{\U{2124} }%
%BeginExpansion
\mathbb{Z}
%EndExpansion
_{\infty}$ and $%
%TCIMACRO{\U{2102} }%
%BeginExpansion
\mathbb{C}
%EndExpansion
^{3}/%
%TCIMACRO{\U{2124} }%
%BeginExpansion
\mathbb{Z}
%EndExpansion
_{\infty}\times%
%TCIMACRO{\U{2124} }%
%BeginExpansion
\mathbb{Z}
%EndExpansion
_{\infty}$ quiver theories\footnote{In the case of $sl_{\infty}$, this is
immediate. \ In the case of $\widehat{sl}\left(  \infty\right)  $, we use the
fact that the specialized character of the basic representation coincides (in
the limit of large central charge) with the partition function of crystal
melting. \ Because the Weyl group of $\widehat{sl}\left(  \infty\right)  $
generates the weight system for the basic representation, we conclude that
each weight appears with multiplicity one, and that correspondingly, the orbit
of the highest weight under the action of the Weyl group generates the
representation.}, we conclude that the 2d and 3d inverted empty room black
holes correspond to highest weight states of the basic representations of
$sl_{\infty}$ and $\widehat{sl}\left(  \infty\right)  $, respectively. \ We
now discuss the physical origin of these algebras.

Although it is well known that the orbifold theory $%
%TCIMACRO{\U{2102} }%
%BeginExpansion
\mathbb{C}
%EndExpansion
^{2}/%
%TCIMACRO{\U{2124} }%
%BeginExpansion
\mathbb{Z}
%EndExpansion
_{n}$ produces an enhanced $SU(n)$ gauge symmetry in the uncompactified
directions of the space-time, it is less clear whether a similar enhanced
gauge symmetry is produced by the orbifold $%
%TCIMACRO{\U{2102} }%
%BeginExpansion
\mathbb{C}
%EndExpansion
^{3}/%
%TCIMACRO{\U{2124} }%
%BeginExpansion
\mathbb{Z}
%EndExpansion
_{n}\times%
%TCIMACRO{\U{2124} }%
%BeginExpansion
\mathbb{Z}
%EndExpansion
_{n}$. \ Nevertheless, the similarities between the algebras $sl_{\infty}$ and
$\widehat{sl}\left(  \infty\right)  $ and their analogous roles in describing
crystal melting suggest that at large $n$ the orbifold $%
%TCIMACRO{\U{2102} }%
%BeginExpansion
\mathbb{C}
%EndExpansion
^{3}/%
%TCIMACRO{\U{2124} }%
%BeginExpansion
\mathbb{Z}
%EndExpansion
_{n}\times%
%TCIMACRO{\U{2124} }%
%BeginExpansion
\mathbb{Z}
%EndExpansion
_{n}$ produces the more exotic affine gauge symmetry $\widehat{SU}(n)$ in the
limit of large central charge. \ In this language, the sum over ranks records
the diagonal $U(1)$ charge remaining after resolving the geometry.

\subsection{2d Crystals and $sl_{\infty}$}

We now show that the 2d inverted tableau charge $Q_{inv}$ of equation
(\ref{invertedTab}) is the highest weight state of the basic representation of
$sl_{\infty}$. \ Some background on the representation theory of $sl_{\infty}$
is collected in appendix C. \ Let $\left\{  \alpha_{i}\right\}  _{i\in%
%TCIMACRO{\U{2124} }%
%BeginExpansion
\mathbb{Z}
%EndExpansion
}$ denote the simple roots and $\left\{  \omega_{i}\right\}  _{i\in%
%TCIMACRO{\U{2124} }%
%BeginExpansion
\mathbb{Z}
%EndExpansion
}$ the fundamental weights of $sl_{\infty}$. \ The relation:
\begin{equation}
\frac{1}{2}Q_{inv}\cdot\alpha_{i}=\delta_{i,0} \label{dotprod}%
\end{equation}
for all $i$ implies the formal identification:%
\begin{equation}
\omega_{0}\sim\frac{1}{2}Q_{inv}. \label{equivdots}%
\end{equation}
The specialized character of the irreducible representation $L(\omega_{0})$ is
\cite{KACTHIRD}:
\begin{equation}
ch_{L(\omega_{0})}\left(  t\rho\right)  =\underset{\mu\in\Omega_{\omega_{0}}%
}{\sum}m_{\mu}\left(  \omega_{0}\right)  e^{\left(  \omega_{0}\cdot
\rho-j\left(  \mu\right)  \right)  t}=q^{-\omega_{0}\cdot\rho}\underset
{n\geq1}{%
%TCIMACRO{\dprod }%
%BeginExpansion
{\displaystyle\prod}
%EndExpansion
}\left(  1-q^{n}\right)  ^{-1} \label{qdimdefinition}%
\end{equation}
where $j(\mu)$ denotes the depth of the weight $\mu$, $q=e^{-t}$, $m_{\mu
}\left(  \omega_{0}\right)  $ $(=1)$ is the multiplicity of $\mu$ and the Weyl
vector $\rho$ is the sum over all the $\omega_{i}$. \ We next compute the
regulated dot product:%
\begin{equation}
\rho\cdot\frac{1}{2}Q_{inv}=\frac{1}{2}\underset{N\rightarrow\infty}{\lim
}\underset{i=-N/2}{\overset{N/2}{\sum}}-\left\vert i\right\vert =-\frac{1}%
{2}\underset{i=1}{\overset{\infty}{\sum}}i=\frac{1}{24}%
\end{equation}
which implies:%
\begin{equation}
ch_{L_{inv}}\left(  q\right)  =q^{-1/24}\underset{n\geq1}{%
%TCIMACRO{\dprod }%
%BeginExpansion
{\displaystyle\prod}
%EndExpansion
}\left(  1-q^{n}\right)  ^{-1}=\eta(q)^{-1} \label{charinverse}%
\end{equation}
where $\eta$ is the Dedekind eta function and we have switched notation to
emphasize the interpretation of $Q_{inv}$ as a highest weight state.
\ Finally, we note that the basic representation of $sl_{\infty}$ is identical
to the chiral boson representation of the Virasoro algebra. \ The details of
the mapping between representations of $sl_{\infty}$ and the Virasoro algebra
may be found in \cite{KACBOMBAY,KACTHIRD}.

\subsection{3d Crystals and $\widehat{sl}\left(  \infty\right)  $}

In the type IIB\ mirror theory, the intersection of the 3d inverted empty room
charge configuration $Q_{inv}$ with the homology 3-cycles $\Delta_{(a,b)}$
which base the quiver theory is:%
\begin{equation}
\frac{1}{3}Q_{inv}\cap\Delta_{(a,b)}=\delta_{(0,0),(a,b)}\text{.}%
\end{equation}
This is the 3d inverted tableau quiver analogue of equation (\ref{dotprod}).

We now show that the 3d inverted empty room charge configuration $Q_{inv}$
defines a highest weight state for a representation of the affine algebra
$\widehat{sl}\left(  \infty\right)  $ with central charge $C\rightarrow\infty
$. \ To this end, we show that the character of the basic representation
reproduces the 3d crystal melting partition sum. \ Because the orbit of the
highest weight state under the Weyl group generates this representation, we
conclude that the inverted empty room black hole is a highest weight state of
$\widehat{sl}\left(  \infty\right)  $ and that the dual charge configurations
are generated by the action of the Weyl/duality group of $\widehat{sl}\left(
\infty\right)  $. \ See appendix D for details on $\widehat{sl}\left(
\infty\right)  $ as well its relation to the $W_{1+\infty}$ algebra.

The unitary representations of this algebra are realized by tensoring $C\geq0$
$bc$ systems with conformal weights $\lambda_{i}+1$ and $-\lambda_{i}$ for the
$i^{th}$ system. \ In the case $\lambda_{i}=\lambda$ for all $i$, the
specialized character of the associated unitary irreducible representation
$L(\lambda,C)$ is \cite{KACW(gl_N),AwataReviewWinf}:%
\begin{equation}
ch_{L(\lambda,C)}=Trq^{L_{0}}=q^{\frac{1}{2}\lambda(\lambda-1)C}\underset
{j=1}{\overset{\infty}{%
%TCIMACRO{\dprod }%
%BeginExpansion
{\displaystyle\prod}
%EndExpansion
}}\underset{k=1}{\overset{C}{%
%TCIMACRO{\dprod }%
%BeginExpansion
{\displaystyle\prod}
%EndExpansion
}}\left(  1-q^{j+k-1}\right)  ^{-1}. \label{charWinf}%
\end{equation}
Expanding this product yields:%
\begin{align}
ch_{L(\lambda,C)}  &  =q^{\frac{1}{2}\lambda(\lambda-1)C}\underset
{j=1}{\overset{\infty}{%
%TCIMACRO{\dprod }%
%BeginExpansion
{\displaystyle\prod}
%EndExpansion
}}\left(  1-q^{j}\right)  ^{-1}\left(  1-q^{j+1}\right)  ^{-1}\cdot\cdot
\cdot\left(  1-q^{j+C-1}\right)  ^{-1}\\
&  =q^{\frac{1}{2}\lambda(\lambda-1)C}\left(  \underset{n=0}{\overset{C}{%
%TCIMACRO{\dsum }%
%BeginExpansion
{\displaystyle\sum}
%EndExpansion
}}p_{3d}(n)q^{n}+O(q^{C+1})\right)
\end{align}
where $p_{3d}(n)$ denotes the number of crystal melting configurations with
$n$ boxes. \ As $C\rightarrow\infty$ the character of $L(\lambda,C)$ tends to
the partition function for 3d crystal melting. \ In fact, as explained in
\cite{KACW(gl_N)}, the states at levels less than $C$ coincide with those of
the quasifinite Verma module given in appendix D by equation
(\ref{VermaStates}).

Roughly speaking, $C$ is an upper bound on the number of 2d tableaux given by
diagonally slicing a three dimensional crystal melting configuration. \ This
implies that in the large $n$ orbifold $%
%TCIMACRO{\U{2102} }%
%BeginExpansion
\mathbb{C}
%EndExpansion
^{3}/%
%TCIMACRO{\U{2124} }%
%BeginExpansion
\mathbb{Z}
%EndExpansion
_{n}\times%
%TCIMACRO{\U{2124} }%
%BeginExpansion
\mathbb{Z}
%EndExpansion
_{m}$, $m\sim C$. \ Indeed, note that when $C=1$ the character is:%
\begin{equation}
ch_{L(\lambda,1)}=q^{\frac{1}{2}\lambda(\lambda-1)}\underset{j=1}%
{\overset{\infty}{%
%TCIMACRO{\dprod }%
%BeginExpansion
{\displaystyle\prod}
%EndExpansion
}}\left(  1-q^{j}\right)  ^{-1}\label{2dtabrecovery}%
\end{equation}
which recovers the partition function for 2d crystal melting. \ Finally, as
observed in \cite{AwatathreeDim}, in the limit $C\rightarrow\infty$ the basic
representation of $\widehat{sl}\left(  \infty\right)  $ is closely related to
the partition function of a three dimensional free field.

\section{Equivariant Sheaves and Topological Strings \label{Topconnection}}

In this section we explain the mathematical connection between our black hole
charge configurations and the crystal melting configurations of topological
string theory. \ We also discuss the sense in which the exceptional collection
defined by the empty room perfect matching is canonically determined by the
geometry $%
%TCIMACRO{\U{2102} }%
%BeginExpansion
\mathbb{C}
%EndExpansion
^{3}/%
%TCIMACRO{\U{2124} }%
%BeginExpansion
\mathbb{Z}
%EndExpansion
_{n}\times%
%TCIMACRO{\U{2124} }%
%BeginExpansion
\mathbb{Z}
%EndExpansion
_{n}$. \ We caution that this material is more formal than other parts of this note.

The A-model partition function on $%
%TCIMACRO{\U{2102} }%
%BeginExpansion
\mathbb{C}
%EndExpansion
^{3}$ coincides with the crystal melting partition function
\cite{CrystalMelting}:%
\begin{equation}
Z_{A}\left(
%TCIMACRO{\U{2102} }%
%BeginExpansion
\mathbb{C}
%EndExpansion
^{3}\right)  =\underset{n\geq1}{%
%TCIMACRO{\dprod }%
%BeginExpansion
{\displaystyle\prod}
%EndExpansion
}\left(  1-q^{n}\right)  ^{-n}=Z_{\text{crystal}}%
\end{equation}
where $q=e^{-g_{s}}$. \ This is also the partition function of the six
dimensional topologically twisted $U(1)$ gauge theory given by a D6-brane
filling $%
%TCIMACRO{\U{2102} }%
%BeginExpansion
\mathbb{C}
%EndExpansion
^{3}$ \cite{QuantumFoam}. \ Mathematically, the instanton configurations of
the D6-brane are specified by ideal sheaves\footnote{An ideal sheaf
corresponds to a torsion free sheaf with vanishing first Chern class. \ For
open sets $U$ in a variety $X$, we define a collection of ideals
$\mathcal{I}(U)\subset\mathcal{O}_{X}$. \ These local data define the
corresponding ideal sheaf $\mathcal{I}$.}. \ Physically, these are singular
gauge field configurations which have vanishing D4-charge and D2- and
D0-charge given respectively by the second and third Chern characters of the
gauge bundle. \ The mathematical theory which counts these ideal sheaves is
known as Donaldson-Thomas theory \cite{DonaldsonThomas}. \ It has recently
been shown that for toric Calabi-Yau threefolds, the Gromov-Witten and
Donaldson-Thomas invariants coincide \cite{GWDTone,GWDTtwo}. \ It is believed
that for more general Calabi-Yau threefolds the partition function of this six
dimensional $U(1)$ gauge theory agrees with the result from Donaldson-Thomas
theory. \ 

Because $%
%TCIMACRO{\U{2102} }%
%BeginExpansion
\mathbb{C}
%EndExpansion
^{3}$ is topologically trivial, the singular gauge field configurations of the
$U(1)$ gauge theory are specified by ideals $I$ generated by monomials in the
ring $%
%TCIMACRO{\U{2102} }%
%BeginExpansion
\mathbb{C}
%EndExpansion
\left[  x,y,z\right]  $. \ Each such ideal determines a collection of points
in $%
%TCIMACRO{\U{2124} }%
%BeginExpansion
\mathbb{Z}
%EndExpansion
_{\geq0}^{3}$:%
\begin{equation}
\pi_{I}=\left\{  \left(  i,j,k\right)  \in%
%TCIMACRO{\U{2124} }%
%BeginExpansion
\mathbb{Z}
%EndExpansion
_{\geq0}^{3}|i,j,k\geq1,\text{ }x^{i-1}y^{j-1}z^{k-1}\notin I\right\}  .
\label{partition}%
\end{equation}
The D4- and D2-brane charge of such a configuration vanishes and the D0-brane
charge is given by the total number of points in $\pi_{I}$ \cite{QuantumFoam}.

In the rest of this section we explain how these ideals are generated from the
perspective of the gauged quiver quantum mechanics. \ Our strategy will be to
exploit the dual meaning of perfect matchings in the two systems. \ On the one
hand, such perfect matchings parametrize crystal melting configurations, and
hence ideal sheaves of $%
%TCIMACRO{\U{2102} }%
%BeginExpansion
\mathbb{C}
%EndExpansion
^{3}$. \ On the other hand, these same perfect matchings parametrize
exceptional collections of sheaves with support on some complex surface in a
resolution of the large $n$ orbifold $%
%TCIMACRO{\U{2102} }%
%BeginExpansion
\mathbb{C}
%EndExpansion
^{3}/%
%TCIMACRO{\U{2124} }%
%BeginExpansion
\mathbb{Z}
%EndExpansion
_{n}\times%
%TCIMACRO{\U{2124} }%
%BeginExpansion
\mathbb{Z}
%EndExpansion
_{n}$. \ In the case of crystal melting configurations, this complex surface
is defined by the canonical resolution of the orbifold singularity.

We first explain the connection between $\Gamma$-equivariant sheaves of $%
%TCIMACRO{\U{2102} }%
%BeginExpansion
\mathbb{C}
%EndExpansion
^{3}$ and tautological sheaves in orbifolds of the form $%
%TCIMACRO{\U{2102} }%
%BeginExpansion
\mathbb{C}
%EndExpansion
^{3}/%
%TCIMACRO{\U{2124} }%
%BeginExpansion
\mathbb{Z}
%EndExpansion
_{n}\times%
%TCIMACRO{\U{2124} }%
%BeginExpansion
\mathbb{Z}
%EndExpansion
_{n}\equiv%
%TCIMACRO{\U{2102} }%
%BeginExpansion
\mathbb{C}
%EndExpansion
^{3}/\Gamma\equiv X$ for arbitrary $n$. \ Following the discussion of the
generalized McKay correspondence in \cite{ReidMcKayReview}, for each
$\rho:\Gamma\rightarrow End\left(  V_{\rho}\right)  $ an irreducible
representation of $\Gamma$, the eigensheaf $F_{\rho}^{\prime}$ on $X$ is:%
\begin{equation}
F_{\rho}^{\prime}\equiv\text{Hom}\left(  V_{\rho},\mathcal{O}_{%
%TCIMACRO{\U{2102} }%
%BeginExpansion
\mathbb{C}
%EndExpansion
^{3}}\right)  ^{\Gamma}.
\end{equation}
The tautological sheaf $F_{\rho}$ has support on the resolution $f:\tilde
{X}\rightarrow X$:%
\begin{equation}
F_{\rho}=f^{\ast}F_{\rho}^{\prime}/torsion.
\end{equation}
The generalized McKay Correspondence of \cite{BKR} now implies that the
$F_{\rho}$ form a basis for the K-theory $K_{0}(\tilde{X})$ and lift to a
basis for the $\Gamma$-equivariant K-theory $K_{0}^{\Gamma}\left(
%TCIMACRO{\U{2102} }%
%BeginExpansion
\mathbb{C}
%EndExpansion
^{3}\right)  $.

Although there are many crepant\footnote{A crepant resolution of a singular
Calabi-Yau $X$ is a smooth resolution $Y$ such that $c_{1}\left(  Y\right)
=0$.} resolutions of the orbifold $%
%TCIMACRO{\U{2102} }%
%BeginExpansion
\mathbb{C}
%EndExpansion
^{3}/\Gamma$ which are all related by flops, there is one distinguished choice
such that the tautological sheaves of the resolution are given by a collection
of line bundles which are generated by their global sections and such that any
positive linear combination of the associated divisors is ample on the
resolution. \ This resolution is a $\Gamma$-equivariant version of the $%
%TCIMACRO{\U{2102} }%
%BeginExpansion
\mathbb{C}
%EndExpansion
^{3}$ Hilbert scheme and is known as $\Gamma$-Hilb$(%
%TCIMACRO{\U{2102} }%
%BeginExpansion
\mathbb{C}
%EndExpansion
^{3})$ in the mathematical literature \cite{GHilbNakamura}.

\begin{figure}
[ptb]
\begin{center}
\includegraphics[
height=1.8049in,
width=5.1855in
]%
{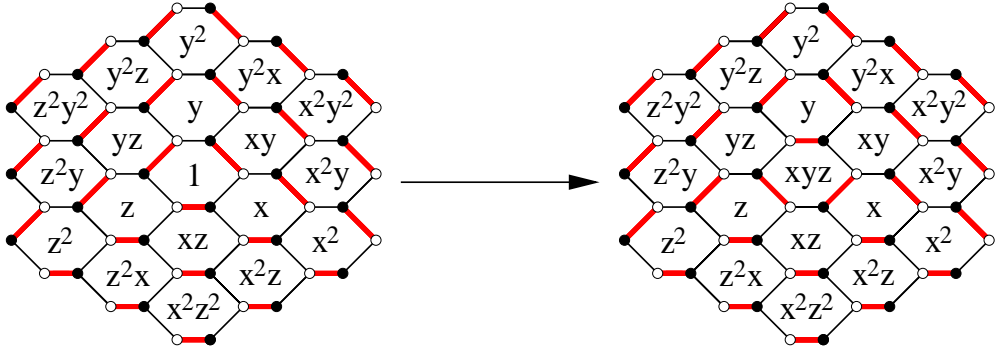}%
\caption{Each sheaf of an exceptional collection supported on the complex
surface defined by the canonical resolution of $\mathbb{C} ^{3}/\mathbb{Z}
_{n}\times\mathbb{Z} _{n}$ lifts to a monomial in the variables $x,y$ and $z$.
\ The figure shows the effect of changing from the empty room perfect matching
to the single box configuration. \ The generators of the two ideals in
$\mathbb{C} \left[  x,y,z\right]  $ are respectively $1$ and $x,y,z$. }%
\label{monomialemptyonewithatoms}%
\end{center}
\end{figure}

As explained in \cite{ReidMcKayReview} and references therein, the
tautological sheaves of $\Gamma$-Hilb$(%
%TCIMACRO{\U{2102} }%
%BeginExpansion
\mathbb{C}
%EndExpansion
^{3})$ may be viewed as a collection of monomials in the ring $%
%TCIMACRO{\U{2102} }%
%BeginExpansion
\mathbb{C}
%EndExpansion
\left[  x,y,z\right]  $. \ Specializing to the case of the large $n$
orbifolds, we now give a local algorithm for converting an exceptional
collection of sheaves supported on a complex surface in $\Gamma$-Hilb$(%
%TCIMACRO{\U{2102} }%
%BeginExpansion
\mathbb{C}
%EndExpansion
^{3})$ into such a collection of monomials. \ Given two quiver nodes $A$ and
$B$ of a Beilinson quiver connected by a bifundamental $X_{AB}$, the
associated monomials are related by multiplication by $x$:%
\begin{equation}
M_{B}=xM_{A} \label{monomap}%
\end{equation}
with similar conventions for multiplication by $y$ and $z$. \ Note that this
removes the overall ambiguity of $\Gamma$-invariant monomials of the form
$xyz$. \ The degrees of these monomials match the rank assignments of the
gauged quiver quantum mechanics. \ The collection of monomials defined by the
empty room perfect matching corresponds to the same collection of tautological
sheaves of $\Gamma$-Hilb$(%
%TCIMACRO{\U{2102} }%
%BeginExpansion
\mathbb{C}
%EndExpansion
^{3})$ discussed in \cite{ReidMcKayReview}. \ Further, when $F$ is the
starting node of the perfect matching $PM_{0}$ with associated monomial
$M_{F}$, the local dimer move which maps $F$ to a terminal node sends $M_{F}$
to the monomial:%
\begin{equation}
M_{F}\rightarrow xyzM_{F}. \label{monorearrange}%
\end{equation}
See figure (\ref{monomialemptyonewithatoms}) for the change from the empty
room perfect matching to the single box perfect matching. \ Similar tilings by
monomials have appeared in the mathematical literature on the McKay
correspondence. \ See \cite{ReidMcKayReview} and references therein for more details.

These monomials generate an ideal $I_{PM}$ in $%
%TCIMACRO{\U{2102} }%
%BeginExpansion
\mathbb{C}
%EndExpansion
\left[  x,y,z\right]  $. \ Equation (\ref{monorearrange}) implies that a local
dimer rearrangement in the monomials translates to removing a point from the
partition $\pi_{I_{PM}}$ defined by equation (\ref{partition}).\ \ Hence,
(with suitable asymptotics) the exceptional collections of the infinite
orbifold lift to monomial generators for ideal sheaves in $%
%TCIMACRO{\U{2102} }%
%BeginExpansion
\mathbb{C}
%EndExpansion
^{3}$.%
%TCIMACRO{\FRAME{ftbpFU}{5.1855in}{1.8049in}{0pt}{\Qcb{Each sheaf of an
%exceptional collection supported on the complex surface defined by the
%canonical resolution of $\U{2102} ^{3}/\U{2124} _{n}\times\U{2124} _{n}$ lifts
%to a monomial in the variables $x,y$ and $z$. \ The figure shows the effect of
%changing from the empty room perfect matching to the single box configuration.
%\ The generators of the two ideals in $\U{2102} \left[  x,y,z\right]  $ are
%respectively $1$ and $x,y,z$. }}{\Qlb{monomialemptyonewithatoms}%
%}{monomialemptyonewithatoms.eps}{\special{ language "Scientific Word";
%type "GRAPHIC";  maintain-aspect-ratio TRUE;  display "USEDEF";
%valid_file "F";  width 5.1855in;  height 1.8049in;  depth 0pt;
%original-width 7.1722in;  original-height 2.4782in;  cropleft "0";
%filename 'monomialemptyonewithatoms.eps';file-properties "XNPEU";}}}%
%BeginExpansion
%EndExpansion

\subsection{2d Analogue}

The partition function for two dimensional crystal melting is given by the
topologically twisted $U(1)$ gauge theory of a D4-brane filling $%
%TCIMACRO{\U{2102} }%
%BeginExpansion
\mathbb{C}
%EndExpansion
^{2}$:%
\begin{equation}
Z_{D4}\left(
%TCIMACRO{\U{2102} }%
%BeginExpansion
\mathbb{C}
%EndExpansion
^{2}\right)  =q^{-1/24}\underset{n\geq1}{%
%TCIMACRO{\dprod }%
%BeginExpansion
{\displaystyle\prod}
%EndExpansion
}\left(  1-q^{n}\right)  ^{-1}.
\end{equation}
Note that the extra factor of $q^{-1/24}$ agrees with the character formula of
equation (\ref{charinverse}).

By resolving $%
%TCIMACRO{\U{2102} }%
%BeginExpansion
\mathbb{C}
%EndExpansion
^{2}/%
%TCIMACRO{\U{2124} }%
%BeginExpansion
\mathbb{Z}
%EndExpansion
_{n}$ we obtain a tautological sheaf for each quiver node. \ Each of these
sheaves lifts to a monomial in $%
%TCIMACRO{\U{2102} }%
%BeginExpansion
\mathbb{C}
%EndExpansion
\left[  x,y\right]  $. \ The analogue of the Beilinson quiver in two
dimensions is given by deleting a minimal number of arrows from the quiver so
that no directed loops remain. \ In this case, the starting (resp. ending)
nodes of the quiver have all incoming (resp. outgoing) bifundamentals deleted.
\ When the monomials $M_{i}$ and $M_{i+1}$ are connected by the arrow
$X_{i,i+1}$ $($resp. $Y_{i+1,i})$, the analogue of equation (\ref{monomap}) is
$M_{i+1}=xM_{i}$ $($resp. $M_{i}=yM_{i+1})$. \ Given a 2d Beilinson quiver
with starting node $i$, the analogue of equation (\ref{monorearrange}) is:%
\begin{equation}
M_{i}\rightarrow xyM_{i}.
\end{equation}

\section{Conclusions}

In this note we have discovered a one to one correspondence between two and
three dimensional crystal melting configurations and certain type
IIA\ BPS\ black holes obtained from wrapping branes on collapsed cycles of the
large $n$ orbifolds $%
%TCIMACRO{\U{2102} }%
%BeginExpansion
\mathbb{C}
%EndExpansion
^{2}/%
%TCIMACRO{\U{2124} }%
%BeginExpansion
\mathbb{Z}
%EndExpansion
_{n}$ and $%
%TCIMACRO{\U{2102} }%
%BeginExpansion
\mathbb{C}
%EndExpansion
^{3}/%
%TCIMACRO{\U{2124} }%
%BeginExpansion
\mathbb{Z}
%EndExpansion
_{n}\times%
%TCIMACRO{\U{2124} }%
%BeginExpansion
\mathbb{Z}
%EndExpansion
_{n}$. \ Moreover, the entire BPS\ spectrum of such black holes is generated
by geometric transitions of brane configurations which leave the classical
background geometry invariant. \ In the process of establishing this
connection, we found a more general set of results on the physical meaning of
perfect matchings in gauge theory dimer models and interpreted the associated
black hole partition functions in terms of the representation theory of
algebras naturally associated to the duality groups of the orbifolds.
\ Finally, we provided a mathematical connection between these black hole
crystal melting configurations and topological string theory. \ In the rest of
this section we speculate on some possible extensions of this work.

Returning to the partition functions of section \ref{PartitionFunctions},
consider the statistical mechanical average value of quiver ranks in the limit
of uniform weighting given by $q\rightarrow1$. \ By suitably rescaling the
ranks and locations of quiver nodes, we obtain a limit shape for the profile
of the molten crystal
\cite{CerfKenyonLimitShape,OkounkovReshetikhin,KOSDimerReview} which defines
the charges of an average BPS\ black hole. \ In the immediate vicinity of a
quiver node all of its neighboring ranks are equal. \ It is therefore tempting
to deconstruct these quiver theories as in \cite{DeconLittle}. \ In our case,
however, we can only deconstruct finite sized patches of the quiver. \ Lifting
to M-theory, this suggests that after many flops a deconstructed M2-brane
emerges with the gradients in the ranks signalling local changes in the
worldvolume curvature.

Our analysis only treated flop transitions which leave the classical geometry
probed by the branes unchanged. \ It would be interesting to see whether more
general transitions of the orbifold theories also have an interpretation in
topological string theory.

Finally, we only considered the simplest manifestation of three dimensional
crystal melting given by the $%
%TCIMACRO{\U{2102} }%
%BeginExpansion
\mathbb{C}
%EndExpansion
^{3}$ Calabi-Yau crystal. \ As a sketch of what to expect for more general
toric Calabi-Yau threefolds, we note that the mirror curve of the conifold
$\mathcal{O}(-1)\oplus\mathcal{O}(-1)\rightarrow\mathbb{P}^{1}$ is described
by dimer models on the infinite square lattice
\cite{CrystalMelting,KOSDimerReview}. \ Auspiciously, this is the same as the
gauge theory dimer of the $%
%TCIMACRO{\U{2124} }%
%BeginExpansion
\mathbb{Z}
%EndExpansion
_{n}\times%
%TCIMACRO{\U{2124} }%
%BeginExpansion
\mathbb{Z}
%EndExpansion
_{n}$ orbifold of the conifold in the large $n$ limit.

\section*{Acknowledgements}

We thank D. Vegh for many helpful discussions and collaboration at an early
stage of this work. \ In addition, we thank S. Franco, C. Herzog, S. Katz, L.
Motl and M. Wijnholt for helpful discussions. \ Part of this work was
performed at the 2006 Simons Workshop in Mathematics and Physics and we thank
the organizers for providing a productive and stimulating environment. \ CV
also thanks the CTP\ at MIT for hospitality during his sabbatical leave. \ The
work of JJH\ and CV is supported in part by NSF grants PHY-0244821 and
DMS-0244464. \ The work of JJH\ is also supported by an NSF\ Graduate Fellowship.

\appendix{}

\section*{Appendix A: Exceptional Collections and Fractional Branes
\label{excepreview}}

This appendix reviews the procedure for obtaining a physical basis of
fractional branes on a non-compact Calabi-Yau threefold $X$ from an
exceptional collection of sheaves supported on a possibly singular complex
surface $V$ given by partially resolving $X$. \ A good reference for
additional information on exceptional collections is \cite{RudakovHelices}.
\ An exceptional collection of sheaves $\mathcal{E}^{V}=\left(  E_{1}%
^{V},...,E_{S}^{V}\right)  $ supported on $V$ is an ordered collection of
sheaves defined by the conditions:%
\begin{align}
\text{Ext}^{q}\left(  E_{i}^{V},E_{i}^{V}\right)   &  =\left\{
\begin{array}
[c]{c}%
0\text{ if }q>0\\%
%TCIMACRO{\U{2102} }%
%BeginExpansion
\mathbb{C}
%EndExpansion
\text{ if }q=0
\end{array}
\right\}  \\
\text{Ext}^{q}\left(  E_{i}^{V},E_{j}^{V}\right)   &  =0\text{ if }i>j
\end{align}
where Ext$^{(\cdot)}$ is a generalization of cohomology for sheaves. \ When an
exceptional collection satisfies the further property that for all $q>0$,%
\begin{equation}
\text{Ext}^{q}\left(  E_{i}^{V},E_{j}^{V}\right)  =0\text{ for all }i\neq j
\end{equation}
it is called strong. \ We define a helix of sheaves $\left\{  E_{i}%
^{V}\right\}  _{i\in%
%TCIMACRO{\U{2124} }%
%BeginExpansion
\mathbb{Z}
%EndExpansion
}$ of period $S$ recursively as follows:%
\begin{align}
E_{i+S}^{V} &  =R_{E_{i+S-1}^{V}}...R_{E_{i+1}^{V}}E_{i}^{V}\label{rightmut}\\
E_{-i}^{V} &  =L_{E_{-i+1}^{V}}...L_{E_{n-1-i}^{V}}E_{n-i}^{V}\text{
\ \ }(i\geq0).\label{period}%
\end{align}
The notation $L_{E}$ and $R_{E}$ denotes respectively left and right mutations
by the sheaf $E$. \ See pages 5 and 6 of \cite{RudakovHelices} for the
definition of mutation for sheaves. \ Given two sheaves $(E,F)$ which form an
exceptional pair, these mutations correspond to braiding operations which
produce another exceptional pair of sheaves:%
\begin{align}
&  (E,F)\overset{L}{\rightarrow}(L_{E}F,E)\\
&  (E,F)\overset{R}{\rightarrow}(F,R_{F}E).
\end{align}
The left and right braiding operations are inverse to one another.

We refer to any exceptional collection which generates $\left\{  E_{i}%
^{V}\right\}  _{i\in%
%TCIMACRO{\U{2124} }%
%BeginExpansion
\mathbb{Z}
%EndExpansion
}$ as a foundation of the helix. \ It follows from the definitions given above
that for every integer $m,$ the collection of sheaves $(E_{m+1}^{V}%
,...,E_{m+S}^{V})$ is also a foundation of the helix $\left\{  E_{i}%
^{V}\right\}  _{i\in%
%TCIMACRO{\U{2124} }%
%BeginExpansion
\mathbb{Z}
%EndExpansion
}$. \ When a helix has a strong foundation it is called a strong helix. \ An
important caveat is that there are multiple orderings of the sheaves inside of
the helix which obey the same periodicity properties. \ In this sense, the
indexing by the integers should be viewed as only a partial
ordering\footnote{This point was emphasized to us by C. Herzog.}.

As proposed in \cite{KontsDerived} and further substantiated in
\cite{DouglasCategories,DouglasFiolSpectrum,DiaconescusDouglasStringyCalabi},
a B-brane is given by a bounded\ complex of coherent sheaves in $D^{b}\left(
X\right)  $. \ Following the discussion in \cite{HerzogSeiberg}, given a
strong exceptional collection of sheaves supported on $V$ which generates the
derived category of coherent sheaves on $V$, the corresponding basis of
fractional branes in $D^{b}(X)$ is given by the collection of left-mutated
objects lifted from $D^{b}\left(  V\right)  $ to $D^{b}\left(  X\right)  $:%
\begin{equation}
\mathcal{E}_{Frac}=\left(  L_{\delta E_{1}^{V}}...L_{\delta E_{S-1}^{V}}\delta
E_{S}^{V},...,L_{\delta E_{1}^{V}}\delta E_{2}^{V},\delta E_{1}^{V}\right)
\equiv\left(  E_{S},..,,E_{1}\right)  \label{fracbasisdef}%
\end{equation}
where the notation $\delta E$ denotes the complex:%
\begin{equation}
...\rightarrow0\rightarrow0\rightarrow E\rightarrow0\rightarrow0\rightarrow...
\end{equation}
with $E$ a coherent sheaf sitting at the $0^{th}$ position of the complex.
\ See page 65 of \cite{RudakovHelices} for the definition of mutation for
objects in the derived category. \ The two collections are dual in the sense
that:%
\begin{equation}
\chi\left(  E_{i}^{V},E_{j}\right)  =\delta_{i,j} \label{orthogbasis}%
\end{equation}
where the pairing $\chi\left(  A,B\right)  $ counts with signs the massless
open string modes between the branes $A$ and $B$. \ Each strong foundation for
a helix defines a physical collection of fractional branes on $X$ which base
the quiver \cite{HerzogSeiberg,AspinwallTilting}.

The upper triangular matrix $S_{ij}=\chi\left(  E_{j},E_{i}\right)  $ with
ones on the diagonal determines the adjacency of bifundamentals in the quiver
gauge theory. \ For $i<j$, $S_{ij}$ is the number of arrows from node $j$ to
node $i$ minus the number from $i$ to $j$. \ In the associated Landau-Ginzburg
theory, $S_{ij}$ corresponds to the soliton counting matrix
\cite{IqbalVafaBranesandMirror}. \ At the level of charges, each $E_{i}$ maps
to a homology 3-sphere $\Delta_{i}$ in the mirror theory. \ Our sign
convention is that for $i<j$:%
\begin{equation}
S_{ij}=\chi\left(  E_{j},E_{i}\right)  =\Delta_{j}\cap\Delta_{i}.
\label{adjacency}%
\end{equation}
Next consider two B-branes $E_{i},E_{j}$ of $\mathcal{E}_{Frac}$ as in
equation (\ref{fracbasisdef}) with mirror 3-cycles $\Delta_{i}$ and
$\Delta_{j}$, respectively. \ For $i<j$, the Chern character of $R_{E_{i}%
}E_{j}$ is\footnote{Note that the ordering of B-branes in equation
(\ref{fracbasisdef}) means that when $i<j$, $E_{i}$ appears after $E_{j}$ in
the collection of fractional branes.}:%
\begin{equation}
ch\left(  R_{E_{i}}E_{j}\right)  =ch\left(  E_{j}\right)  -\chi\left(
E_{j},E_{i}\right)  ch\left(  E_{i}\right)  =ch\left(  E_{j}\right)
-S_{ij}ch\left(  E_{i}\right)  .
\end{equation}
This maps to the mirror 3-cycle:%
\begin{equation}
ch\left(  R_{E_{i}}E_{j}\right)  \rightarrow\Delta_{j}+\left(  \Delta_{i}%
\cap\Delta_{j}\right)  \Delta_{i}. \label{mirrorrighty}%
\end{equation}
Similarly, the Chern character of $L_{E_{j}}E_{i}$ is:%
\begin{equation}
ch\left(  L_{E_{j}}E_{i}\right)  =ch\left(  E_{i}\right)  -\chi\left(
E_{j},E_{i}\right)  ch\left(  E_{j}\right)  =ch\left(  E_{i}\right)
-S_{ij}ch\left(  E_{j}\right)
\end{equation}
which maps to the mirror 3-cycle:%
\begin{equation}
ch\left(  L_{E_{j}}E_{i}\right)  \rightarrow\Delta_{i}+\left(  \Delta_{i}%
\cap\Delta_{j}\right)  \Delta_{j}=\Delta_{i}-\left(  \Delta_{j}\cap\Delta
_{i}\right)  \Delta_{j}.
\end{equation}
Comparing with the discussion in subsection \ref{hom}, we see that an
appropriate combination of right (resp. left) mutations will realize the
transformation $S_{R}^{2}$ (resp. $S_{L}^{2}$) in the mirror theory.

\section*{Appendix B: Dimer Moves and Flops \label{DimermoveFlop}}

In this appendix we establish the link between geometry preserving flops and
local rearrangements of perfect matchings. \ Each internal perfect matching of
a gauge theory dimer model defines an exceptional collection of sheaves
supported on a complex surface obtained from a partial resolution of the toric
Calabi-Yau threefold $X$ \cite{HHVExceptional}. \ As shown in
\cite{HananyKennaway}, these internal perfect matchings also label the
internal grid points of the toric diagram for $X$. \ When two perfect
matchings correspond to the same internal grid point, they determine different
foundations of the same helix. \ We now show that a local rearrangement of an
internal perfect matching corresponds to a series of right (resp. left)
mutations of the starting (resp. terminal) sheaf of the Beilinson quiver. \ In
the mirror type IIB\ theory, this corresponds to the transformation $S_{R}%
^{2}$ (resp. $S_{L}^{2}$).

First fix an internal perfect matching $PM$ with associated strong exceptional
collection of sheaves $\mathcal{E}^{PM}=(E_{1}^{PM},...,E_{S}^{PM})$ supported
on the complex surface $V_{PM}$. \ With conventions as in appendix A, the
sheaf $E_{1}^{PM}$ corresponds to a face $F$ in the dimer model such that all
of the incoming arrows of $F$ belong to the perfect matching. \ Next consider
the internal perfect matching $PM^{\prime}$ such that the formal difference of
edges $PM-PM^{\prime}$ forms a closed loop encircling the face $F$. \ This
defines another exceptional collection $\mathcal{E}^{PM^{\prime}}=\left(
E_{1}^{PM^{\prime}},...,E_{S}^{PM^{\prime}}\right)  $ with support on $V_{PM}%
$. \ We assume that the corresponding rearrangement of perfect matchings only
alters the sheaf $E_{1}^{PM}$ of the collection $\mathcal{E}^{PM}$ so that:%
\begin{equation}
E_{i-1}^{PM^{\prime}}=E_{i}^{PM}%
\end{equation}
for $i=2,...,S$. \ To determine the effect on the sheaf $E_{1}^{PM}$, note
that the helix condition implies the collection $\left(  E_{2}^{PM}%
,...,E_{S}^{PM},E_{S+1}^{PM}\right)  $ is strongly exceptional. \ Since all of
the outgoing arrows of the sheaf $E_{S+1}^{PM}$ are absent from the associated
Beilinson quiver, we conclude that\footnote{We thank C. Herzog for
correspondence on this point.}:%
\begin{equation}
E_{S}^{PM^{\prime}}=E_{S+1}^{PM}=R_{E_{S}^{PM}}...R_{E_{2}^{PM}}E_{1}^{PM}
\label{StartFin}%
\end{equation}
where the second equality follows from equation (\ref{rightmut}).

Next consider the effect of a local dimer move on a terminal node of the
Beilinson quiver. \ In this case, an argument similar to the one given above
implies that the exceptional collection of sheaves changes to:%
\begin{equation}
E_{i+1}^{PM^{\prime}}=E_{i}^{PM}%
\end{equation}
for $i=1,...,S-1$ and further,%
\begin{equation}
E_{1}^{PM^{\prime}}=E_{0}^{PM}=L_{E_{1}^{PM}}\cdot\cdot\cdot L_{E_{S-1}^{PM}%
}E_{S}^{PM}. \label{FinStart}%
\end{equation}

In the event that there are two possible starting (resp. terminal) nodes for
the Beilinson quiver, there are then two distinct orderings of the sheaves in
the exceptional collection. \ The two local dimer rearrangements correspond to
right (resp. left) mutating the chosen starting (resp. terminal) sheaf to the
right (resp. left) of all other sheaves in the collection.

We now demonstrate that the local dimer moves considered above correspond to
the transformations $S_{R}^{2}$ and $S_{L}^{2}$ in the mirror theory. \ Given
a quiver node $i$, we split the remaining quiver nodes into three types: those
that are outgoing from $i$, those that are incoming to $i$ and those that do
not touch $i$. \ We label the sheaves of the exceptional collection according
to this convention as well. \ In the case $i=1$, we have outgoing sheaves
$E_{2}^{PM},...,E_{a}^{PM}$ and incoming sheaves $E_{a+1}^{PM},...,E_{S}^{PM}%
$, where by abuse of notation we have labelled all nodes which do not touch
$i$ as incoming. \ Now right mutate $E_{1}$ through the outgoing sheaves. \ We
refer to this \textquotedblleft helix duality\textquotedblright\ as $H_{R}$.
\ This transformation produces another exceptional collection:%
\begin{equation}
H_{R}\left(  \mathcal{E}^{PM}\right)  =\left(  E_{2}^{PM},...,E_{a}%
^{PM},R_{E_{a}^{PM}}\cdot\cdot\cdot R_{E_{2}^{PM}}E_{1}^{PM},E_{a+1}%
^{PM},...,E_{S}^{PM}\right)  .
\end{equation}
The corresponding transformation on the basis of fractional branes was
computed in \cite{HerzogSeiberg} with the result:%
\begin{equation}
H_{R}\left(  \mathcal{E}_{Frac}\right)  =\left(  E_{S},...,E_{a+1},\delta
E_{1}^{PM}\left[  1\right]  ,R_{\delta E_{1}^{PM}}E_{a},...,R_{\delta
E_{1}^{PM}}E_{2}\right)  \label{fracSeibone}%
\end{equation}
where the the object $\mathcal{F}\left[  n\right]  $ in~$D^{b}\left(
X\right)  $ denotes the complex $\mathcal{F}$ with all entries shifted $n$
positions to the left. \ Note that the complex corresponding to the dualized
node has shifted one position to the left. \ This has the effect of exchanging
the brane for the anti-brane and thus reverses the direction of all incoming
and outgoing arrows incident on the corresponding quiver node. \ Performing
another $H_{R}$ duality therefore right mutates the sheaf $R_{E_{a}^{PM}}%
\cdot\cdot\cdot R_{E_{2}^{PM}}E_{1}^{PM}$ through the sheaves $E_{a+1}%
^{PM},...,E_{S}^{PM}$:%
\begin{equation}
H_{R}^{2}\left(  \mathcal{E}^{PM}\right)  =\left(  E_{2}^{PM},...,E_{S}%
^{PM},R_{E_{S}^{PM}}\cdot\cdot\cdot R_{E_{2}^{PM}}E_{1}^{PM}\right)  .
\label{rightdouble}%
\end{equation}
The corresponding transformation on the basis of fractional branes is:%
\begin{equation}
H_{R}^{2}\left(  \mathcal{E}_{Frac}\right)  =\left(  \delta E_{1}^{PM}\left[
2\right]  ,R_{\delta E_{1}^{PM}[1]}E_{S},...,R_{\delta E_{1}^{PM}[1]}%
E_{a+1},R_{\delta E_{1}^{PM}}E_{a},...,R_{\delta E_{1}^{PM}}E_{2}\right)  .
\label{finalseibs}%
\end{equation}

Next consider the \textquotedblleft helix duality\textquotedblright\ $H_{L}$
given by left mutating $E_{S}^{PM}$ past all of its incoming sheaves. \ We
label the incoming and outgoing sheaves as $E_{b+1}^{PM},...,E_{S-1}^{PM}$ and
$E_{1}^{PM},...,E_{b}^{PM}$, respectively. \ Applying the transformation
$H_{L}$ yields:%
\begin{equation}
H_{L}\left(  \mathcal{E}^{PM}\right)  =\left(  E_{1}^{PM},...,E_{b}%
^{PM},L_{E_{b+1}^{PM}}\cdot\cdot\cdot L_{E_{S-1}^{PM}}E_{S}^{PM},E_{b+1}%
^{PM},...,E_{S-1}^{PM}\right)  .
\end{equation}
The corresponding transformation on the basis of fractional branes is:%
\begin{equation}
H_{L}\left(  \mathcal{E}_{Frac}\right)  =\left(  L_{\delta E_{S}^{PM}}%
E_{S-1},...,L_{\delta E_{S}^{PM}}E_{b+1},\delta E_{S}^{PM}\left[  -1\right]
,E_{b},...,E_{1}\right)  \text{.}%
\end{equation}
As before, $\delta E_{S}^{PM}$ has been sent to its \textquotedblleft
anti-brane\textquotedblright, although it is now $\delta E_{S}^{PM}\left[
-1\right]  $ rather than $\delta E_{S}^{PM}\left[  1\right]  $. \ Performing
another $H_{L}$ transformation produces the exceptional collection:%
\begin{equation}
H_{L}^{2}\left(  \mathcal{E}^{PM}\right)  =\left(  L_{E_{1}^{PM}}\cdot
\cdot\cdot L_{E_{S-1}^{PM}}E_{S}^{PM},E_{1}^{PM},...,E_{b}^{PM},E_{b+1}%
^{PM},...,E_{S-1}^{PM}\right)  \label{leftdouble}%
\end{equation}
with corresponding basis of fractional branes:%
\begin{equation}
H_{L}^{2}\left(  \mathcal{E}_{Frac}\right)  =\left(  L_{\delta E_{S}^{PM}%
}E_{S-1},...,L_{\delta E_{S}^{PM}}E_{b+1},L_{\delta E_{S}^{PM}\left[
-1\right]  }E_{b},...,L_{\delta E_{S}^{PM}\left[  -1\right]  }E_{1},\delta
E_{S}^{PM}\left[  -2\right]  \right)  \text{.} \label{finalleftseib}%
\end{equation}
We recognize the transformation of equation (\ref{rightdouble}) as the local
rearrangement of $PM$ given by equation (\ref{StartFin}). \ It follows from
equation (\ref{mirrorrighty}) that in terms of homology cycles in the mirror
theory, $H_{R}^{2}$ corresponds to passing the brane wrapping $\Delta_{1}$
through all outgoing and then all incoming branes. \ An analogous argument
shows that the local rearrangement of $PM$ given by equation (\ref{FinStart})
corresponds to passing the brane wrapping $\Delta_{S}$ through all incoming
and then all outgoing branes. \ Similar computations of monodromy
transformations in terms of mutations have appeared in
\cite{MukaiMcKayorbifolds,TomasielloBeilinson,MayrKahlerPhases}. \ As
discussed in \cite{AspinwallNavigation}, however, there is in general a
difference between monodromy transformations and mutations in the derived category.

Finally, note that under the transformation $H_{R}^{2}$ (resp. $H_{L}^{2}$),
the fractional brane corresponding to $\delta E_{1}^{PM}$ $($resp. $\delta
E_{S}^{PM})$ shifts to $\delta E_{1}^{PM}\left[  2\right]  $ $($resp. $\delta
E_{S}^{PM}\left[  -2\right]  )$ and also moves to a different position in the
collection of fractional branes which base the quiver. \ We thus see that it
is too naive to assume as we did in subsection \ref{hom} that the brane
corresponding to the dualized node simply returns to itself.

\section*{Appendix C: Representations of $sl_{\infty}$ \label{slinfrev}}

This appendix reviews material from \cite{KACBOMBAY} and \cite{KACTHIRD} on
the correspondence between states in the basic representation of $sl_{\infty}$
and 2d Young tableaux. $\ $These 2d tableaux also correspond to the states of
the chiral boson representation of the Virasoro algebra.

The algebra $sl_{\infty}$ is defined as the space of traceless infinite
matrices with only finitely many non-zero entries. \ Given an infinite vector
space $V$ with basis $\left\{  v_{i}\right\}  _{i\in%
%TCIMACRO{\U{2124} }%
%BeginExpansion
\mathbb{Z}
%EndExpansion
}$, we define the infinite wedge space $F=\wedge^{\infty}V$ as the complex
vector space spanned by \textquotedblleft semi-infinite
monomials\textquotedblright:%
\begin{equation}
v_{i_{1}}\wedge v_{i_{2}}\wedge...
\end{equation}
where $i_{1}>i_{2}>...,$ and $i_{n}=i_{n-1}-1$ for $n>>0$. \ $F$ defines a
representation $r$ of $sl_{\infty}$:%
\begin{equation}
r(a)\left(  v_{i_{1}}\wedge v_{i_{2}}\wedge...\right)  =\left(  a\cdot
v_{i_{1}}\right)  \wedge v_{i_{2}}\wedge...+v_{i_{1}}\wedge\left(  a\cdot
v_{i_{2}}\right)  \wedge...+...
\end{equation}
where $a\cdot v$ denotes matrix multiplication of the vector $v\in V$ by the
matrix $a\in sl_{\infty}$. \ For each integer $m\in%
%TCIMACRO{\U{2124} }%
%BeginExpansion
\mathbb{Z}
%EndExpansion
$, we define a charge $m$ vacuum vector:%
\begin{equation}
\left\vert m\right\rangle \equiv v_{m}\wedge v_{m-1}\wedge v_{m-2}\wedge...
\end{equation}
and define the \textquotedblleft charge-m\textquotedblright\ subspace
$F^{(m)}$ as the linear span of all semi-infinite monomials which differ from
$\left\vert m\right\rangle $ in only a finite number of places. \ $F$
decomposes into independent charge spaces:%
\begin{equation}
F=\underset{m\in%
%TCIMACRO{\U{2124} }%
%BeginExpansion
\mathbb{Z}
%EndExpansion
}{\oplus}F^{(m)}. \label{chargedecomp}%
\end{equation}
Each $F^{(m)}$ defines an irreducible representation of $sl_{\infty}$
isomorphic to the basic representation $L\left(  \omega_{m}\right)  $, where
$\omega_{m}$ denotes the fundamental weight such that $\omega_{m}\cdot
\alpha_{n}=\delta_{m,n}$ for all simple roots $\alpha_{n}$. \ The
correspondence between the generators of $F^{(m)}$ and 2d partitions of the
form $\left\{  \lambda_{1}\geq\lambda_{2}\geq\cdot\cdot\cdot\geq0\right\}  $
is given by the bijection:%
\begin{equation}
v_{i_{1}}\wedge v_{i_{2}}\wedge...\mapsto\left\{  \lambda_{1}=i_{1}%
-m,\lambda_{2}=i_{2}-(m-1),...\right\}  .
\end{equation}
Because the $\lambda_{i}$ correspond to the lengths of rows in a 2d Young
tableau, we obtain the expected correspondence.

\section*{Appendix D:\ Representations of $\widehat{sl}\left(  \infty\right)
$ and $W_{1+\infty}$ \label{WinfReprev}}

This appendix reviews the representation theory of the algebras $\widehat
{sl}\left(  \infty\right)  $ and $W_{1+\infty}$. \ Our discussion closely
follows that in \cite{AwataReviewWinf} where further details may be found.
\ The algebra $\widetilde{gl}\left(  \infty\right)  $ consists of infinite
matrices with only a finite number of non-zero diagonals. \ Note that this is
a much larger space than $gl_{\infty}$ which consists of infinite matrices
with only a finite number of non-zero entries. \ The algebra $\widehat
{gl}\left(  \infty\right)  $\ is a central extension of $\widetilde{gl}\left(
\infty\right)  $ and is spanned by generators $E(r,s)$ for $r,s\in%
%TCIMACRO{\U{2124} }%
%BeginExpansion
\mathbb{Z}
%EndExpansion
$ subject to the commutation relations:%
\begin{equation}
\left[  E\left(  r,s),E(r^{\prime},s^{\prime}\right)  \right]  =\delta
_{r^{\prime}+s,0}E(r,s^{\prime})-\delta_{r+s^{\prime},0}E(r^{\prime
},s)+C\delta_{r+s^{\prime},0}\delta_{r^{\prime}+s,0}\left(  \theta_{r}%
-\theta_{r^{\prime}}\right)
\end{equation}
where $\theta_{r}$ equals $1$ for $r\geq0$ and $0$ otherwise. \ 

The $W_{1+\infty}$ algebra is generated by polynomials of $z,$ $z^{-1}$ and
$D\equiv z\partial_{z}$ subject to the commutation relations:%
\begin{equation}
\left[  W\left(  z^{n}e^{xD}\right)  ,W\left(  z^{m}e^{yD}\right)  \right]
=(e^{mx}-e^{ny})W(z^{n+m}e^{(x+y)D})-C\frac{e^{mx}-e^{ny}}{e^{x+y}-1}%
\delta_{n+m,0}%
\end{equation}
where $n,m\in%
%TCIMACRO{\U{2124} }%
%BeginExpansion
\mathbb{Z}
%EndExpansion
$, $C$ is the central charge of the algebra and$~x$ and $y$ are place keeping
devices in the expansion of the exponentials. \ The Cartan subalgebra is
generated by the operators $W(D^{k})$ for $k\geq0$. \ The generators
$L_{n}\equiv-W\left(  z^{n}D\right)  $ span a Virasoro subalgebra with central
charge $c_{VIR}=-2C$. \ The realization of the $W_{1+\infty}$ algebra in terms
of $\widehat{gl}\left(  \infty\right)  $ is given by:%
\begin{equation}
W\left(  z^{n}e^{xD}\right)  =\underset{r+s=n}{\underset{r,s\in%
%TCIMACRO{\U{2124} }%
%BeginExpansion
\mathbb{Z}
%EndExpansion
}{\sum}}e^{x\left(  \lambda-s\right)  }E(r,s)-C\frac{e^{\lambda x}-1}{e^{x}%
-1}\delta_{n,0} \label{Wrealizationingl}%
\end{equation}
where $\lambda$ is a real parameter which we will eventually identify with the
conformal weight of a field in the free field realization of the algebra.
\ Although infinite, the above sum is restricted to the $n^{th}$ diagonal.

A highest weight representation of $W_{1+\infty}$ is determined by a highest
weight state $\left\vert \lambda\right\rangle $ such that:%
\begin{align}
W\left(  z^{n}D^{k}\right)  \left\vert \lambda\right\rangle  &  =0\text{
}(n\geq1,k\geq0)\\
W(D^{k})\left\vert \lambda\right\rangle  &  =\Delta_{k}\left\vert
\lambda\right\rangle \text{ \ }(k\geq0).
\end{align}
In general, the weights $\Delta_{k}$ may be complex numbers which we encode as
coefficients in the weight function:%
\begin{equation}
\Delta\left(  x\right)  =-\underset{k\geq0}{%
%TCIMACRO{\dsum }%
%BeginExpansion
{\displaystyle\sum}
%EndExpansion
}\Delta_{k}\frac{x^{k}}{k!}.
\end{equation}
The Verma module associated with $\left\vert \lambda\right\rangle $ is given
by all descendants of the form:%
\begin{equation}
W\left(  z^{-n_{1}}D^{k_{1}}\right)  ...W\left(  z^{-n_{m}}D^{k_{m}}\right)
\left\vert \lambda\right\rangle
\end{equation}
where $n_{i}\geq1$ and $k_{i}\geq0$ for all $i$. \ From the perspective of the
Virasoro subalgebra, the level of the given descendant is $n_{1}+...+n_{m}$.
\ We obtain an irreducible representation of $W_{1+\infty}$ by projecting out
the null states of the above module. \ In contrast to representations of the
Virasoro algebra, note that there are an infinite number of states at each
level. \ In spite of this, there exist representations of $W_{1+\infty}$ which
have a finite number of independent states at each level \cite{KacRadul}.
\ Such representations are called quasifinite.

A highest weight state $\left\vert \lambda\right\rangle $ of $\widehat
{gl}\left(  \infty\right)  $ is defined by the conditions:%
\begin{align}
E\left(  r,s\right)  \left\vert \lambda\right\rangle  &  =0\text{
\ \ \ \ \ }(r+s>0)\\
E(r,-r)\left\vert \lambda\right\rangle  &  =q_{r}\left\vert \lambda
\right\rangle \text{ \ \ \ \ }(r\in%
%TCIMACRO{\U{2124} }%
%BeginExpansion
\mathbb{Z}
%EndExpansion
).
\end{align}
A representation of $\widehat{gl}(\infty)$ is quasifinite only when a finite
number of $h_{r}=q_{r}-q_{r-1}+C\delta_{r,0}$ are different from zero
\cite{KacRadul}. \ The descendants of $\left\vert \lambda\right\rangle $ are
of the form:%
\begin{equation}
E\left(  -r_{1},-s_{1}\right)  \cdot\cdot\cdot E\left(  -r_{n},-s_{n}\right)
\left\vert \lambda\right\rangle \label{descent}%
\end{equation}
where $r\geq0$ and $s\geq1$. \ The commutation relation:%
\begin{equation}
\left[  W\left(  D^{1}\right)  ,E\left(  r,s\right)  \right]  =\left(
r+s\right)  E\left(  r,s\right)
\end{equation}
implies that the state of equation (\ref{descent}) is an eigenstate of
$W\left(  D^{1}\right)  $ with eigenvalue $-r_{1}-s_{1}-r_{2}-s_{2}%
-...-r_{n}-s_{n}$. \ In fact, one of the key results of \cite{KacRadul}
establishes that the quasifinite representations of $\widehat{gl}\left(
\infty\right)  $ and $W_{1+\infty}$ are identical.

The quasifinite Verma module $V\left(  \lambda,C\right)  $ of the highest
weight state with weight function:%
\begin{equation}
\Delta\left(  x\right)  =C\frac{e^{\lambda x}-1}{e^{x}-1}%
\end{equation}
is spanned by states of the form:%
\begin{equation}
W\left(  z^{-n}D^{k}\right)  \left\vert \mu\right\rangle \label{VermaStates}%
\end{equation}
where $0\leq k\leq n-1$ and $n\geq0$ and $\left\vert \mu\right\rangle $
corresponds to a descendant in the module. \ The specialized character of the
quasifinite Verma module is:%
\begin{equation}
ch_{Verma}=Trq^{L_{0}}=q^{\frac{1}{2}\lambda(\lambda-1)C}\underset{n\geq1}{%
%TCIMACRO{\dprod }%
%BeginExpansion
{\displaystyle\prod}
%EndExpansion
}\left(  1-q^{n}\right)  ^{-n}. \label{Verma}%
\end{equation}
When $C$ is not an integer, the quasifinite Verma module is in fact an
irreducible representation of $W_{1+\infty}$. \ Unfortunately, representations
with non-integer $C$ are not unitary, making the physical interpretation
unclear. \ In the case of the unitary representations, the above Verma module
will contain null vectors which must be projected out. \ Necessary and
sufficient conditions for a unitary representation were found in
\cite{KacRadul} and are given by $C\in%
%TCIMACRO{\U{2124} }%
%BeginExpansion
\mathbb{Z}
%EndExpansion
_{\geq0}$ and:%
\begin{equation}
\Delta\left(  x\right)  =\underset{i=1}{\overset{C}{\sum}}\frac{e^{\lambda
_{i}x}-1}{e^{x}-1}%
\end{equation}
where the $\lambda_{i}$ are real numbers. \ We note that such a representation
is obtained from the tensor product of $C$ $bc$ systems with associated
conformal weights $\lambda_{i}+1$ and $-\lambda_{i}$ for $1\leq i\leq C$.

\bibliographystyle{ssg}
\bibliography{CrystalMeltingBlackHoles}

\end{document}